\documentclass[10pt,twocolumn]{IEEEtran}

\usepackage{amsmath,amsfonts,graphics,amssymb,epsfig,subfigure,color,cite}                             
\usepackage{amsthm,textcomp}
\theoremstyle{plain}
\newtheoremstyle{mystyle}
  {0mm}
  {0mm}
  {}
  {4mm}
  {\bfseries}
  {:}
  { }
  {\thmname{#1}\thmnumber{ #2}\thmnote{ (#3)}}
\theoremstyle{mystyle}
\usepackage{bm}
\usepackage{array}
\usepackage{multirow}
\usepackage{enumerate}
\usepackage{algorithm}
\usepackage{algpseudocode}
\usepackage{adjustbox}
\usepackage{algcompatible}
\algblockdefx{FORP}{ENDFORP}[1]%
  {\textbf{for }#1 \textbf{do in parallel}}%
  {\textbf{end}}
\algblockdefx{FOR}{ENDFOR}[1]%
  {\textbf{for }#1 \textbf{do}}%
  {\textbf{end for}}
\algblockdefx{noEnd}{noEndEnd}[1]
    {#1}
    {{\Statex}}
\algblockdefx{BULLET}{EndBULLET}{$\bullet$ }{}
\algnotext{EndBULLET}
\algblockdefx{IF}{ENDIF}[1]%
  {\textbf{if }#1 \textbf{then}}%
  {\textbf{end if}}
\algnewcommand\algorithmicprocedure{\textbf{function}}
\algnewcommand\FUNC{\item[\algorithmicprocedure]}%
\algnewcommand\algorithmicendprocedure{\textbf{end function}}
\algnewcommand\ENDFUNC{\item[\algorithmicendprocedure]}%
 \usepackage{setspace}
\let\Algorithm\algorithm
\renewcommand\algorithm[1][]{\Algorithm[#1]\setstretch{1.4}}
\usepackage{float}
\usepackage{makecell}
\usepackage{graphicx}
\usepackage{mathrsfs}
\usepackage[caption=false,font=normalsize,labelfont=sf,textfont=sf]{subfig}
\usepackage[right=0.625in, left=0.625in, top=0.7in, bottom=1.0in]{geometry}
\usepackage{verbatim}

\usepackage{bbm}

\makeatletter
\newcommand{\vast}{\bBigg@{4.5}}
\newcommand{\Vast}{\bBigg@{7.5}}
\makeatother

\allowdisplaybreaks

\begin{document}
\title{Cooperative Multi-Satellite ISAC Networks: Centralized vs. Distributed Sensing}
\author{Jeongbin Kim, Jaehong Jo, Seunghyeon Jeon, Yo-Seb Jeon, \IEEEmembership{Member,~IEEE}, \\
Wonjae Shin, \IEEEmembership{Senior Member,~IEEE},   and H. Vincent Poor, \IEEEmembership{Life Fellow,~IEEE}
\vspace{-0.3cm}
\thanks{Jeongbin Kim, Jaehong Jo, Seunghyeon Jeon, and Yo-Seb Jeon are with the Department of Electrical Engineering, POSTECH, Pohang, Gyeongbuk 37673, South Korea (e-mail: jb2067@postech.ac.kr; jaehongjo@postech.ac.kr; seunghyeon.jeon@postech.ac.kr; yoseb.jeon@postech.ac.kr).}
\thanks{Wonjae Shin is with the School of Electrical Engineering, Korea University, Seoul 02841, South Korea (email: wjshin@korea.ac.kr).}
\thanks{H. Vincent Poor is with the Department of Electrical Engineering, Princeton University, Princeton, NJ 08544 (e-mail: poor@princeton.edu).}
\vspace{-0.5cm}
}


\maketitle

\begin{abstract} 
This paper investigates a downlink multi-satellite integrated sensing and communication (ISAC) network, in which multiple satellites simultaneously transmit ISAC signals to provide communication services to ground user equipments and enable cooperative sensing of airborne targets through multiple gateways. To support this dual functionality, we introduce communication and sensing beamforming designs based on uniform planar arrays with optimized power allocation. Building on these designs, we propose two cooperative sensing frameworks, namely centralized and distributed. In the centralized framework, each gateway forwards its sensing observations to a central unit (CU), where the positions of multiple targets are jointly estimated from the aggregated data. To mitigate the signaling overhead inherent in centralized processing, a distributed framework is further proposed, in which each gateway independently estimates target positions and transmits only the local estimates to the CU. To associate estimates from different gateways, a data association problem based on the squared Euclidean distance is formulated and efficiently solved using the Hungarian algorithm. The final target positions are then obtained by minimizing the distance error. Simulation results demonstrate that the proposed centralized and distributed frameworks significantly outperform existing sensing schemes while satisfying communication performance requirements. \textcolor{black}{The results further show that, for the grid-free algorithms, the centralized framework achieves higher sensing accuracy than the distributed framework by fully exploiting the aggregated observations. Conversely, for the grid-based algorithms, the distributed framework outperforms the centralized one under sufficient sensing resources, owing to its ability to resolve off-grid targets through data fusion.} Finally, we evaluate the sensing-communication trade-off from the viewpoints of sensing accuracy and communication power consumption under the proposed frameworks.
\end{abstract}

\begin{IEEEkeywords}
Integrated sensing and communication (ISAC), cooperative sensing, centralized sensing, distributed sensing,  multi-satellite network.
\end{IEEEkeywords}

\section{Introduction}\label{Sec:Intro}
Satellite systems have attracted considerable attention as a key technology in 5G and beyond-5G wireless networks for supporting ubiquitous and global connectivity \cite{kodheli2020satellite}, \cite{jeon2025beam}.
This evolution has been further supported by the standardization efforts of the 3rd Generation Partnership Project (3GPP), which has incorporated non-terrestrial networks (NTNs) into enhanced mobile broadband and machine-type communication frameworks \cite{darwish2022leo}. Nevertheless, the overall system performance of an individual satellite is inherently limited by constrained onboard power budgets and processing capabilities, which hinder its ability to meet the demands of NTNs in terms of throughput, latency, and reliability. To overcome these bottlenecks, cooperative satellite systems have been introduced, which are expected to provide low-latency and ultra-reliable communication services with seamless coverage via inter-satellite cooperation \cite{he2024spatial}. Recently, the deployment of low Earth orbit (LEO) satellites has accelerated the growth of dense and large-scale satellite networks \cite{del2019technical}. Building upon this infrastructure, LEO satellites have expanded their utility beyond connectivity to support a wide range of services, including Earth monitoring, positioning, and navigation \cite{liu2021leo}. However, the advancement of LEO satellite systems toward dual roles in communication and sensing has inevitably introduced new challenges associated with orbital and spectral constraints, such as spectrum congestion, inter-satellite interference, and complex constellation design \cite{yin2024integrated}. 

Integrated sensing and communication (ISAC) has been considered a promising technology for future wireless networks due to its unified signal processing and common hardware for communication and sensing. This integration simultaneously provides wireless communication and sensing within a shared spectrum and thus achieves reduced hardware cost, improved spectral efficiency, and mitigated spectrum conflicts \cite{wei2023integrated}. In a cell-free multiple-input multiple-output (MIMO) system, multiple ISAC transceivers can collaborate to perform multistatic sensing and jointly serve communication users for enhancing sensing robustness and communication reliability over wide areas \cite{shi2022device}. Motivated by this, multi-satellite ISAC networks have been studied to facilitate high-accuracy sensing and reliable communication with global coverage. Nevertheless, multi-satellite ISAC networks still face technical challenges for practical realization, which mainly lie in the design of cooperation frameworks.

The cooperative ISAC networks can be categorized into two types based on their cooperation level, namely centralized and distributed frameworks \cite{meng2024cooperative}. The centralized framework jointly processes all observations from the spatially distributed sensor network. In contrast, the distributed framework consists of weak cooperation, in which each sensor independently performs local sensing and shares its results for subsequent fusion to improve accuracy. While centralized frameworks can achieve high sensing accuracy by fully exploiting global information, they generally require considerable fronthaul capacity and computational resources at a central unit (CU), which may limit their scalability and practical deployment. On the other hand, distributed frameworks significantly reduce fronthaul overhead and computational burden by relying on local processing, at the cost of potentially degraded sensing performance due to limited information exchange. This trade-off between sensing accuracy, communication overhead, and implementation complexity motivates a systematic investigation of cooperative frameworks in multi-satellite ISAC networks.

\subsection{Related Works}

\subsubsection{\textcolor{black}{Cooperative Multi-Sensor Localization}}
\textcolor{black}{There has been extensive prior work on cooperative localization with spatially distributed sensors. The authors in \cite{subedi2016group} proposed a multi-target tracking method that jointly exploits Doppler measurements collected from multiple sensors. This cooperation provides complementary information that cannot be obtained from a single-sensor Doppler measurement. The authors in \cite{shen2020group} proposed a cooperative localization method to estimate the direction of arrival (DoA) of far-field sources and localize near-field sources. These studies employed group sparse reconstruction for centralized localization. However, these works do not consider beamforming design or its integration with ISAC. Moreover, they focus exclusively on centralized localization and therefore do not provide a systematic comparison between centralized and distributed frameworks.}

\subsubsection{Cooperative Terrestrial ISAC}
Most existing studies on cooperative sensing have been conducted in terrestrial scenarios. For example, the studies in \cite{yang2025cooperative}, \cite{nozari2025toward} investigated centralized sensing frameworks, where observations collected from multiple nodes equipped with uniform linear arrays are jointly processed at a CU for target detection.
Specifically, the target detection problem was solved via sparse Bayesian learning \cite{yang2025cooperative} and iterative node selection and refinement \cite{nozari2025toward}. Other studies investigated distributed sensing frameworks, where each sensing node independently processes its received reflected signals and forwards the extracted information, such as angle of arrival (AoA) pairs \cite{lyu2024irs} or range information \cite{shi2022device}, \cite{shi2024joint}, to a CU for joint estimation via data fusion. The authors in \cite{lou2024beamforming} proposed both centralized and distributed frameworks by considering different backhaul capacity constraints and corresponding information fusion strategies. \textcolor{black}{From a network-level perspective, the authors in \cite{meng2024scaling} characterized the ergodic communication rate and the scaling law of the localization Cramér-Rao lower bound (CRLB) using stochastic geometry, based on which the cooperative cluster sizes and transmit power were optimized to balance the two functionalities.} However, these works primarily target terrestrial sensing scenarios and are not directly applicable to multi-satellite ISAC networks due to the fundamentally different deployment geometry.

\subsubsection{Non-Cooperative Single-Satellite ISAC}
Most existing studies on satellite ISAC networks have focused on non-cooperative single-satellite scenarios. 
In \cite{you2022beam}, a beam squint-aware hybrid beamforming design was proposed for a massive MIMO system to maximize energy efficiency while guaranteeing sensing performance. 
Similarly, the authors in \cite{kong2025cooperative} investigated ISAC beamforming design in satellite-terrestrial networks by capitalizing on the complementary characteristics of satellite and terrestrial networks.
The authors in \cite{park2025bistatic} proposed a rate-splitting multiple access-based precoder that jointly considers communication rate and the CRLB of AoA estimation.
In addition, this study presented an AoA estimation method based on the multiple signal classification (MUSIC) algorithm. The authors in \cite{liu2025efficient} designed pilot allocation to balance the trade-off between communication and navigation for orthogonal frequency division multiple access-based delay and Doppler estimation. 
Unfortunately, the sensing performance of the single-satellite ISAC network is fundamentally limited by the lack of spatial diversity. This limitation highlights the need for cooperative multi-satellite ISAC networks.

\subsubsection{Cooperative Multi-Satellite ISAC}
Only a few studies have explored cooperative sensing frameworks tailored to multi-satellite ISAC networks.
The total CRLB for time difference of arrival estimation in multi-target scenarios was optimized by designing beam hopping and power allocation \cite{wang2021cooperative}, and beam scheduling and transmit beamforming \cite{xv2023joint}. These studies focused solely on beamforming design without considering a sensing framework. Another study \cite{wang2025multiple} investigated communication and sensing beamforming to minimize the CRLB for angle of departure (AoD) and reflection coefficient estimation. This study also proposed a particle swarm optimization (PSO)-based centralized location sensing algorithm. 
Despite these early research efforts, none of the existing works has systematically investigated a multi-target sensing process, which is essential for extending the applicability of satellite ISAC networks. Moreover, an effective distributed sensing framework for multi-satellite ISAC networks remains largely unexplored, despite its potential to overcome the high signaling overhead and limited scalability inherent in centralized cooperation for sensing.

\subsection{Contributions}
To fill this research gap, in this paper, we present cooperative multi-target sensing frameworks for multi-satellite ISAC networks, an aspect not investigated in existing studies. In this setup, multiple satellites and gateways are deployed to simultaneously provide communication services to ground user equipments (UEs) and perform airborne target sensing. To this end, we design the communication and sensing beamforming based on the uniform planar array (UPA) structure. For practicality, a hybrid analog-digital beamforming architecture is also considered to reduce power consumption and hardware cost for
large antenna arrays. Subsequently, we develop both centralized and distributed sensing frameworks to provide comprehensive insights into cooperation design. We also present an analysis of the two frameworks from the viewpoints of fronthaul overhead and computational complexity. Through extensive simulations, we not only compare the performance of the proposed centralized and distributed frameworks, but also validate the superiority of these frameworks over conventional benchmark schemes. The key contributions of this paper are summarized as follows:
\begin{itemize}
\item We introduce novel cooperative sensing frameworks for multi-satellite ISAC networks, where multiple satellites transmit ISAC signals to provide communication services to ground UEs and perform target sensing, while gateways receive reflected signals from the targets for cooperative sensing. For this dual-purpose operation, we propose a hybrid beamforming design that jointly supports sensing and communication, aiming to maximize the sensing signal-to-interference-plus-noise ratio (SINR) while minimizing the communication power under quality-of-service (QoS) constraints.
\item We propose a centralized sensing framework, where multiple gateways cooperatively perform sensing by transmitting their local observations to a CU. Based on these aggregated observations, \textcolor{black}{we formulate the multi-target sensing task and develop cooperative sensing algorithms for efficient target localization.
The proposed algorithms encompass two search strategies, which are referred to as grid-based and grid-free methods.}
\item We also propose a distributed framework, where each gateway individually estimates the target positions and forwards its results to the CU. To match the estimates associated with the same target, we perform data association by sequentially applying the Hungarian algorithm. After the association, the target positions are determined by solving a distance error minimization problem with a closed-form solution.
\end{itemize}

\textcolor{black}{This work extends our previous study \cite{conference}, which focused only on a distributed sensing framework with a grid-based sensing algorithm. In this work, we develop a centralized framework for multi-satellite ISAC networks and introduce grid-free sensing algorithms for both frameworks. The proposed frameworks are evaluated via extensive simulations under various environments and compared with benchmark schemes as well as our prior work. This extension provides deeper insights into cooperation design in multi-satellite ISAC networks.}

\textit{Notations:} 
Scalars, vectors, and matrices are denoted by italic, lowercase boldface, and uppercase boldface letters, respectively. 
For a vector \(\mathbf{x}\), \(\mathbf{x}^T\), \(\mathbf{x}^H\), \([\mathbf{x}]_n\), and \(|\mathbf{x}|_p\) denote its transpose, Hermitian transpose, \(n\)-th element, and \(\ell_p\)-norm, respectively. For a matrix \(\mathbf{X}\), \(\mathbf{X}^{-1}\), \([\mathbf{X}]{:,n}\), and \([\mathbf{X}]_{m,n}\) denote its inverse, \(n\)-th column, and \((m,n)\)-th entry. Moreover, \(\otimes\), \(|\cdot|\), \(\operatorname{blkdiag}(\cdot)\), \(\operatorname{Re}(\cdot)\), and \(\operatorname{Im}(\cdot)\) have their conventional meanings. We use \(j=\sqrt{-1}\), \(\mathbf{0}_N\), and \(\mathbf{I}_N\) for the imaginary unit, zero vector, and identity matrix, respectively. Finally, \(|\mathcal{S}|\) denotes set cardinality, and \(\mathcal{CN}(\boldsymbol{\mu},\boldsymbol{\Sigma})\) denotes a complex Gaussian distribution with mean \(\boldsymbol{\mu}\) and covariance \(\boldsymbol{\Sigma}\).

\begin{figure}[t]
  \centering
   \includegraphics[scale=0.95,width=0.95\linewidth]{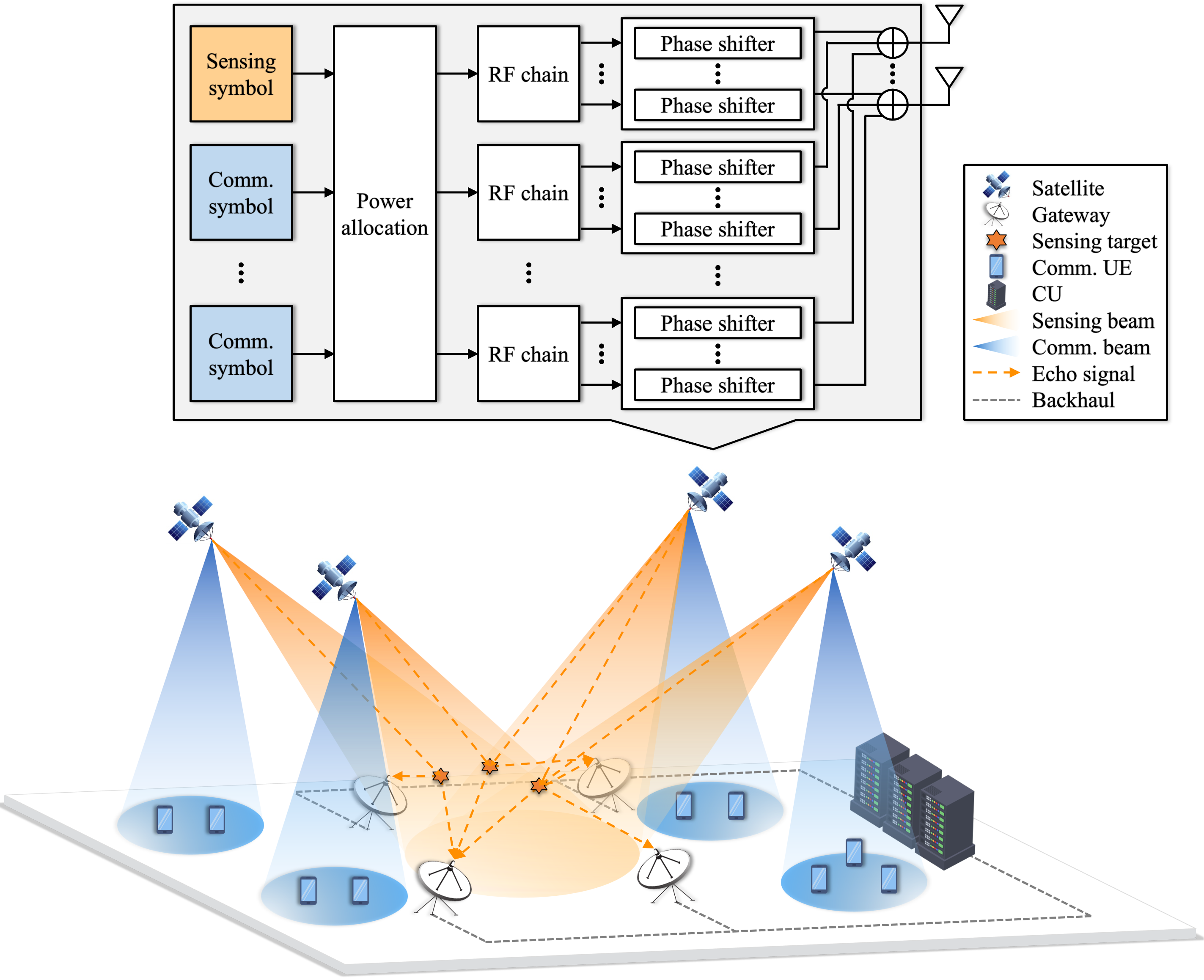}
    \caption{System model of the proposed multi-satellite ISAC network.}
    \label{Fig_1}
    \vspace{-3mm}
\end{figure}

\section{System Model}\label{Sec:System}
Consider a downlink multi-satellite ISAC network, where $I$ satellites and $L$ gateways are deployed to simultaneously provide communication services to ground UEs and perform passive target sensing over $T$ time slots, as illustrated in Fig. \ref{Fig_1}. Each satellite is equipped with a half-wavelength spaced UPA. The arrays consist of $N^{\textrm{sat}} = N_{x}^{\textrm{sat}} \times N_{y}^{\textrm{sat}}$ antenna elements, where $N_{x}^{\textrm{sat}}$ and $N_{y}^{\textrm{sat}}$ are the numbers of elements along the $x$- and $y$-axes, respectively. Similarly, each gateway is equipped with the same type of UPA, comprising $N^{\textrm{gat}} = N_{x}^{\textrm{gat}} \times N_{y}^{\textrm{gat}}$ elements. Satellite $i$ transmits ISAC signals to detect $K$ airborne targets and communicate with $U_i$ single-antenna UEs. \textcolor{black}{Meanwhile, gateways do not transmit signals but receive reflected echo signals from the targets for cooperative sensing. The considered LEO-transmission and ground-reception architecture is particularly suitable for large-scale monitoring of airborne targets. Unlike monostatic satellite sensing, this architecture does not require satellites to receive weak target echoes onboard. Instead, the reflected signals are collected by gateways, which can be densely deployed and equipped with large receive arrays. These characteristics make the proposed architecture attractive for wide-area airborne target monitoring.} We assume that UEs are randomly and uniformly distributed within the communication footprint of the serving satellite. The communication footprint is assumed to be non-overlapping with those of other satellites. In contrast, the sensing region is shared by all satellites to enable joint target sensing and is assumed to be disjoint from any communication footprint. The sets of satellites, gateways, UEs served by satellite $i$, and targets are denoted by $\mathcal{I}=\{1,\cdots,I\}$, $\mathcal{L}=\{1,\cdots,L\}$, $\mathcal{U}_i=\{1,\cdots,U_i\}$, and $\mathcal{K}=\{1,\cdots,K\}$, respectively. We also denote the fixed 3D positions of satellite $i$, gateway $l$, the $u$-th UE served by satellite $i$, and target $k$ as $\mathbf{q}^{\textrm{sat}}_{i} = [x^{\textrm{sat}}_{i},  y^{\textrm{sat}}_{i}, z^{\textrm{sat}}_{i}]^T$, $\mathbf{q}^{\textrm{gat}}_{l} = [x^{\textrm{gat}}_{l},  y^{\textrm{gat}}_{l}, z^{\textrm{gat}}_{l}]^T$, $\mathbf{q}^{\textrm{ue}}_{i,u} = [x^{\textrm{ue}}_{i,u},  y^{\textrm{ue}}_{i,u}, z^{\textrm{ue}}_{i,u}]^T$, and $\mathbf{q}^{\textrm{tar}}_{k} = [x^{\textrm{tar}}_{k},  y^{\textrm{tar}}_{k}, z^{\textrm{tar}}_{k}]^T $, respectively. \textcolor{black}{It is assumed that satellites and targets are static over the considered time interval. This is because their displacements are negligible due to the short duration of this interval.\footnote{\textcolor{black}{For example, the channel bandwidth is $5 \textrm{ MHz}$ according to 3GPP R1-2401845 \cite{3gppR1_2401845}. In this case, the duration of each slot is $0.2 \textrm{ $\mu$s}$. For a LEO speed of $7.5 \textrm{ km/s}$ \cite{yoo2024cache} and $T=324$ time slots, the satellite displacement is $0.486 \textrm{ m}$, which is sufficiently small for the static assumption. Similarly, even for a high-speed airborne target such as the Global Hawk with a speed of $203 \textrm{ m/s}$ \cite{karapantazis2005broadband}, the target displacement over the same time interval is only about $0.013 \textrm{ m}$. Thus, range migration and Doppler variation within the considered time interval are negligible.  For this reason, target motion is ignored within each time interval, and target trajectories can be tracked over successive intervals.}}} The position information of satellites, gateways, and UEs is perfectly available to all nodes in the network, whereas the positions of targets remain unknown.

\subsection{Channel Model}\label{Sec:Channel Model}
Typically, the satellite-to-UE channel has a strong line-of-sight (LoS) path and a limited number of weak non-LoS (NLoS) paths due to the high orbital altitude of the satellite. Although the high relative velocity of the satellite induces a Doppler effect, it can be effectively compensated by means of time-frequency synchronization techniques \cite{park2025bistatic}. Motivated by these facts, the communication link between satellite $i$ and the $u$-th UE associated with satellite $j$ can be modeled as a Rician fading channel \cite{yoo2024cache}, which is given by
\begin{align}
\mathbf{h}_{i,j,u}(t) = \frac{\lambda\sqrt{A^{\textrm{sat}}_{\textrm{Tx}}A^{\textrm{ue}}_{\textrm{Rx}}}}{4\pi d^{\textrm{sat-ue}}_{i,j,u}}\bigg(\sqrt{\frac{\kappa}{1+\kappa}}\mathbf{h}^{\textrm{LoS}}_{i,j,u} + \sqrt{\frac{1}{1+\kappa}}\mathbf{h}^{\textrm{NLoS}}_{i,j,u}(t)\bigg),
\end{align}
where $\lambda$ is the carrier wavelength, $A^{\textrm{sat}}_{\textrm{Tx}}$ is the transmit antenna gain of the satellite, $A^{\textrm{ue}}_{\textrm{Rx}}$ is the receive antenna gain of the UE, $d^{\textrm{sat-ue}}_{i,j,u} = \| \mathbf{q}^{\textrm{sat}}_{i}-\mathbf{q}^{\textrm{ue}}_{j,u}\|_{2}$ denotes the distance between satellite $i$ and the $u$-th UE associated with satellite $j$, and $\kappa$ is the Rician factor. Also, $\mathbf{h}^{\textrm{LoS}}_{i,j,u}=e^{-j\frac{2\pi d^{\textrm{sat-ue}}_{i,j,u}}{\lambda}} \mathbf{v}^{\textrm{sat-ue}}_{i,j,u}$ and $\mathbf{h}^{\textrm{NLoS}}_{i,j,u}(t) \sim
\mathcal{CN}(\mathbf{0}_{N^{\textrm{sat}}},\mathbf{I}_{N^{\textrm{sat}}})$ are the LoS and NLoS components, respectively. Here, the array response vector $\mathbf{v}^{\textrm{sat-ue}}_{i,j,u} \in \mathbb{C}^{N^{\textrm{sat}}}$ is constructed based on the AoD of the satellite-to-UE link as $\mathbf{v}^{\textrm{sat-ue}}_{i,j,u} = \mathbf{v}^{\textrm{sat-ue}}_{x,i,j,u}\otimes \mathbf{v}^{\textrm{sat-ue}}_{y,i,j,u}$,
where $\mathbf{v}^{\textrm{sat-ue}}_{x,i,j,u} \in \mathbb{C}^{N_{x}^{\textrm{sat}}}$ and $\mathbf{v}^{\textrm{sat-ue}}_{y,i,j,u} \in \mathbb{C}^{N_{y}^{\textrm{sat}}}$ denote the array responses along the $x$- and $y$-axes, respectively. The $n$-th element of each response is expressed as
\begin{align}
&\big[\mathbf{v}^{\textrm{sat-ue}}_{x,i,j,u}\big]_{n} = e^{-j\pi(n-1)\cos\phi^{\textrm{sat-ue}}_{i,j,u}\cos\theta^{\textrm{sat-ue}}_{i,j,u}} ,
\label{Array response sat-ue x axis}
\end{align}
\begin{align}
&\big[\mathbf{v}^{\textrm{sat-ue}}_{y,i,j,u}\big]_n = e^{-j\pi(n-1)\sin\phi^{\textrm{sat-ue}}_{i,j,u}\cos\theta^{\textrm{sat-ue}}_{i,j,u}},
\label{Array response sat-ue y axis}
\end{align}
with the azimuth angle w.r.t. the negative $y$-axis given by
\begin{align}
\phi^{\textrm{sat-ue}}_{i,j,u} = \tan^{-1}\left(- \frac{x^{\textrm{sat}}_{i}-x^{\textrm{ue}}_{j,u}}{y^{\textrm{sat}}_{i}-y^{\textrm{ue}}_{j,u}} \right),
\end{align}
and the elevation angle w.r.t. the negative $z$-axis given by
\begin{align}
\theta^{\textrm{sat-ue}}_{i,j,u} = \tan^{-1}\left(\frac{\sqrt{(x^{\textrm{sat}}_{i}-x^{\textrm{ue}}_{j,u})^2+(y^{\textrm{sat}}_{i}-y^{\textrm{ue}}_{j,u})^2}}{z^{\textrm{sat}}_{i}-z^{\textrm{ue}}_{j,u}} \right).
\end{align}

Similarly, the satellite-to-target and target-to-gateway channels are assumed to be dominated by a LoS component due to the lack of scatterers. For point-like targets, the sensing channel from satellite $i$ to gateway $l$ via target $k$ can be expressed as
\begin{align}
\mathbf{G}_{i,k,l}=&\rho_{i,k,l} \sqrt{\frac{\lambda^2 A^{\textrm{gat}}_{\textrm{Rx}}A^{\textrm{sat}}_{\textrm{Tx}}}{64\pi^3 (d^{\textrm{gat-tar}}_{l,k})^2(d^{\textrm{sat-tar}}_{i,k})^2}} \nonumber \\ 
&\times e^{-j2\pi\frac{ d^{\textrm{gat-tar}}_{l,k}+d^{\textrm{sat-tar}}_{i,k}}{\lambda}}\mathbf{v}^{\textrm{gat-tar}}_{l,k} (\mathbf{v}^{\textrm{sat-tar}}_{i,k})^H,  
\end{align}
where $\rho_{i,k,l}$ represents the unknown reflection coefficient of target $k$ over the bistatic path between satellite $i$ and gateway $l$. According to the Swerling-I model \cite{richards2005fundamentals}, the complex reflection coefficient $\rho_{i,k,l}$ is assumed to be distributed as $\mathcal{CN}(0, \gamma_{i,k,l})$ with radar cross section $\gamma_{i,k,l}$. \textcolor{black}{The reflection coefficient can also be considered time-invariant within the short time interval.} Here, the receive antenna gain of the gateway is denoted by $A^{\textrm{gat}}_{\textrm{Rx}}$, and the distances from gateway $l$ and satellite $i$ to target $k$ are given by $d^{\textrm{gat-tar}}_{l,k} = \| \mathbf{q}^{\textrm{gat}}_{l}-\mathbf{q}^{\textrm{tar}}_{k}\|_{2}$ and $d^{\textrm{sat-tar}}_{i,k} = \| \mathbf{q}^{\textrm{sat}}_{i}-\mathbf{q}^{\textrm{tar}}_{k}\|_{2}$, respectively. The array response vectors for the links from satellite $i$ to target $k$ and from target $k$ to gateway $l$ are denoted by $\mathbf{v}^{\textrm{sat-tar}}_{i,k} \in \mathbb{C}^{N^{\textrm{sat}}}$ and $\mathbf{v}^{\textrm{gat-tar}}_{l,k}\in \mathbb{C}^{N^{\textrm{gat}}}$, respectively. 
Note that the elevation angle between gateway $l$ and target $k$ is defined w.r.t. the positive $z$-axis as
\begin{align}
\theta^{\textrm{gat-tar}}_{l,k} = \tan^{-1}\left(-\frac{\sqrt{(x^{\textrm{gat}}_{l}-x^{\textrm{tar}}_{k})^2+(y^{\textrm{gat}}_{l}-y^{\textrm{tar}}_{k})^2}}{z^{\textrm{gat}}_{l}-z^{\textrm{tar}}_{k}} \right).
\end{align}

\begin{figure*}[!ht]
\vspace{-5mm}
    \begin{align}
    \textrm{SINR}^{\textrm{ue}}_{i,u}(t) = \frac{p_{i,u}^{\textrm{c}}(t)\left|\mathbf{h}^H_{i,i,u}(t)\mathbf{f}_{i,u}^{\textrm{c}}(t)\right|^2}{\!\sum\limits_{v\in\mathcal{U}_i\setminus\left\{u\right\}}\!p_{i,v}^{\textrm{c}}(t)\left|\mathbf{h}^H_{i,i,u}(t)\mathbf{f}_{i,v}^{\textrm{c}}(t)\right|^2+\!\sum\limits_{j\in\mathcal{I}\setminus\left\{i\right\}}\sum\limits_{v\in\mathcal{U}_j}p_{j,v}^{\textrm{c}}(t)\left|\mathbf{h}^H_{j,i,u}(t)\mathbf{f}_{j,v}^{\textrm{c}}(t)\right|^2+\sum\limits_{j\in\mathcal{I}}p_{j}^{\textrm{r}}(t)\left|\mathbf{h}^H_{j,i,u}(t)\mathbf{f}_{j}^{\textrm{r}}(t)\right|^2+\sigma_{i,u}^2}.
    \label{SINR UE} 
    \tag{10}
    \end{align}
    \hrulefill
\vspace{-4mm}
\end{figure*}

\subsection{Signal Model}\label{Sec:Signal Model}
Satellites transmit communication signals jointly with a dedicated sensing signal using a common waveform. 
The transmitted ISAC signal of satellite $i$ at time $t$ is then given by
\begin{align}
\mathbf{x}_i(t) =  \sqrt{p_{i}^{\textrm{r}}(t)}\mathbf{f}_{i}^{\textrm{r}}(t)s_{i}^{\textrm{r}}(t) + \sum_{u\in\mathcal{U}_i} \sqrt{p_{i,u}^{\textrm{c}}(t)}\mathbf{f}_{i,u}^{\textrm{c}}(t)s_{i,u}^{\textrm{c}}(t),
\end{align}
where $p_i^{\textrm{r}}(t)\geq 0 $, $\mathbf{f}_{i}^{\textrm{r}}(t)\in\mathbb{C}^{N^{\textrm{sat}}}$, and $s_{i}^{\textrm{r}}(t) \in\mathbb{C}$ represent the sensing signal power, the normalized sensing beamforming vector, and the sensing symbol, respectively. Similarly, the signal power, normalized beamforming vector, and communication symbol used for the $u$-th UE associated with satellite $i$ are denoted as $p_{i,u}^{\textrm{c}}(t)\geq 0$, $\mathbf{f}_{i,u}^{\textrm{c}}(t)\in\mathbb{C}^{N^{\textrm{sat}}}$, and $s_{i,u}^{\textrm{c}}(t)\in\mathbb{C}$, respectively. We note that all transmitted communication and sensing symbols are zero-mean, unit power, and mutually independent. 

The received signal of the $u$-th UE associated with satellite $i$ at time $t$ is expressed as
\begin{align}
y_{i,u}^{\textrm{ue}}(t)
&=\underbrace{\sqrt{p_{i,u}^{\textrm{c}}(t)}\mathbf{h}^H_{i,i,u}(t)\mathbf{f}_{i,u}^{\textrm{c}}(t)s_{i,u}^{\textrm{c}}(t)}_{\textrm{Desired signal}}+ \underbrace{n^{\textrm{ue}}_{i,u}(t)}_{\textrm{AWGN}}\nonumber\\
&~~~+ \underbrace{\sum_{j\in\mathcal{I}}\sqrt{p_{j}^{\textrm{r}}(t)}\mathbf{h}^H_{j,i,u}(t)\mathbf{f}_{j}^{\textrm{r}}(t)s_{j}^{\textrm{r}}(t)}_{\textrm{Sensing interference}}\nonumber\\
&~~~+ \underbrace{\sum_{j\in\mathcal{I}\setminus\left\{i\right\}}\sum_{v\in\mathcal{U}_j}\sqrt{p_{j,v}^{\textrm{c}}(t)}\mathbf{h}^{H}_{j,i,u}(t)\mathbf{f}_{j,v}^{\textrm{c}}(t)s_{j,v}^{\textrm{c}}(t)}_{\textrm{Inter-cell interference}}\nonumber\\
&~~~+\underbrace{\sum_{v\in\mathcal{U}_i\setminus\left\{u\right\}}\sqrt{p_{i,v}^{\textrm{c}}(t)}\mathbf{h}^H_{i,i,u}(t)\mathbf{f}_{i,v}^{\textrm{c}}(t)s_{i,v}^{\textrm{c}}(t)}_{\textrm{Intra-cell interference}},
\label{received signal UE}
\end{align}
where $n^{\textrm{ue}}_{i,u}(t)\sim \mathcal{CN}(0,\sigma_{i,u}^2)$ is the additive white Gaussian noise (AWGN). We consider the SINR as the performance metric for communication. From \eqref{received signal UE}, the instantaneous SINR of the $u$-th UE served by satellite $i$ at time $t$ can be obtained as shown in \eqref{SINR UE} (at the top of this page). 

For target sensing, the gateways employ a receive beamforming technique. Then, the echo signal received by gateway $l$ at time $t$ is expressed as
\setcounter{equation}{10}
\begin{align}
y_{l}^{\textrm{gat}}(t) = \sum_{k\in\mathcal{K}}\sum_{i\in\mathcal{I}}  \mathbf{w}^{H}_{l}(t) \mathbf{G}_{i,k,l}\mathbf{x}_{i}(t) + \mathbf{w}^{H}_{l}(t)\mathbf{n}^{\textrm{gat}}_l(t),
\label{received signal gateway}
\end{align}
where $\mathbf{w}_{l}(t)\in\mathbb{C}^{N^{\textrm{gat}}}$ denotes the receive beamforming vector of gateway $l$, and $\mathbf{n}_{l}^{\textrm{gat}}(t)\sim\mathcal{CN}(\mathbf{0}_{N^{\textrm{gat}}},\zeta_l^2\mathbf{I}_{N^{\textrm{gat}}})$ represents the AWGN vector. By defining $a_{i,k,l}^{\textrm{gat}}(t) = \rho^{-1}_{i,k,l} \mathbf{w}^{H}_{l}(t) \mathbf{G}_{i,k,l}\mathbf{x}_{i}(t)$, the echo signal in \eqref{received signal gateway} can be rewritten as
\begin{align}
y_{l}^{\textrm{gat}}(t) = \sum_{k\in\mathcal{K}}\sum_{i\in\mathcal{I}} \rho_{i,k,l}  a_{i,k,l}^{\textrm{gat}}(t) + \mathbf{w}^{H}_{l}(t)\mathbf{n}^{\textrm{gat}}_l(t).
\end{align}

\section{Proposed ISAC Beamforming Design}\label{Sec:ISAC Beamforming Design}
This section presents the beamforming design for the multi-satellite ISAC network. We focus on a hybrid analog-digital beamforming architecture to provide a more feasible implementation for large-scale antenna arrays \cite{molisch2017hybrid}. Specifically, satellite $i$ is assumed to employ $U_i+1$ radio frequency (RF) chains, each connected to $N^{\textrm{sat}}$ phase shifters. Among these, $U_i$ RF chains are allocated for communication, while the remaining RF chain is dedicated to sensing. Similarly, gateway $l$ is equipped with one RF chain to receive the sensing signal. Based on this hybrid architecture, the beamforming design must satisfy the following constraints:
\begin{align}
& |[\mathbf{w}_{l}(t)]_{n}|= \frac{1}{\sqrt{N^{\textrm{gat}}}},~~l\in\mathcal{L},\,t\in\mathcal{T},\,n\in\mathcal{N}^{\textrm{gat}}, \label{hybrid architecture radar receive}\\
&|[\mathbf{f}_{i}^{\textrm{r}}(t)]_{n}| = \frac{1}{\sqrt{N^{\textrm{sat}}}},~~i\in\mathcal{I},\,t\in\mathcal{T},\,n\in\mathcal{N}^{\textrm{sat}},\label{hybrid architecture radar transmit}\\
&|[\mathbf{f}_{i,u}^{\textrm{c}}(t)]_{n}| = \frac{1}{\sqrt{N^{\textrm{sat}}}},~~i\in\mathcal{I},\,u\in\mathcal{U}_i,\, t\in\mathcal{T},\,n\in\mathcal{N}^{\textrm{sat}}, \label{hybrid architecture communication transmit}
\end{align}
where $\mathcal{T}=\{1,\cdots,T\}$, $\mathcal{N}^{\textrm{sat}} = \{1,\cdots,N^{\textrm{sat}}\}$ and $\mathcal{N}^{\textrm{gat}} = \{1,\cdots,N^{\textrm{gat}}\}$. \textcolor{black}{Under these constraints, we first describe a simple grid structure that discretizes the continuous sensing region into grid points. Based on this structure,} we present the proposed designs of the communication and sensing beamforming vectors, respectively. 

\subsection{Grid Structure for Efficient Target Sensing}\label{Sec:Grid Structure for Target Sensing}
We consider a grid structure in which the 3D sensing region is discretized into $M$ grid points denoted as $\mathbf{q}^{\textrm{grid}}_{m} = [x^{\textrm{grid}}_{m},  y^{\textrm{grid}}_{m}, z^{\textrm{grid}}_{m}]^T$ for $m \in \mathcal{M}=\{1,\cdots,M\}$. 
For the $m$-th grid point, let $a^{\textrm{grid}}_{i,m,l}(t)$ denote the beamformed sensing channel from satellite $i$ to gateway $l$ via reflection at the $m$-th grid point at time $t$ without the reflection coefficient. 
\textcolor{black}{The communication symbols are not directly available at the gateways unless they are explicitly conveyed to them.
Then, $a^{\textrm{grid}}_{i,m,l}(t)$ is expressed as 
\begin{align}
a^{\textrm{grid}}_{i,m,l}(t)=&\sqrt{\frac{\lambda^2 A^{\textrm{gat}}_{\textrm{Rx}}A^{\textrm{sat}}_{\textrm{Tx}}p_{i}^{\textrm{r}}(t)}{64\pi^3 (d^{\textrm{gat-grid}}_{l,m})^2(d^{\textrm{sat-grid}}_{i,m})^2}}e^{-j2\pi\frac{ d^{\textrm{gat-grid}}_{l,m}+d^{\textrm{sat-grid}}_{i,m}}{\lambda}}\nonumber\\
&\times \mathbf{w}^{H}_{l}(t)\mathbf{v}^{\textrm{gat-grid}}_{l,m} (\mathbf{v}^{\textrm{sat-grid}}_{i,m})^{H}  \mathbf{f}_{i}^{\textrm{r}}(t)s_{i}^{\textrm{r}}(t),
\label{beamformed radar channel}
\end{align}}
where $d^{\textrm{gat-grid}}_{l,m} = \| \mathbf{q}^{\textrm{gat}}_{l}-\mathbf{q}^{\textrm{grid}}_{m}\|_{2}$ and $d^{\textrm{sat-grid}}_{i,m} = \| \mathbf{q}^{\textrm{sat}}_{i}-\mathbf{q}^{\textrm{grid}}_{m}\|_{2}$ represent the distances from the $m$-th grid point to gateway $l$ and satellite $i$, respectively, and the corresponding array response vectors are denoted by $\mathbf{v}^{\textrm{gat-grid}}_{l,m}$ and $\mathbf{v}^{\textrm{sat-grid}}_{i,m}$, respectively. It is worth mentioning that $a^{\textrm{grid}}_{i,m,l}(t)$ is available at both satellites and gateways. 

\begin{figure*}[!ht]
\vspace{-5mm}
    \begin{align}
    \underbrace{
    \begin{bmatrix}
    \chi_{1,1,1,1}(t)&-\tau_{\textrm{c}} \chi_{1,1,1,2}(t)&\cdots&-\tau_{\textrm{c}} \chi_{1,1,I,U_I}(t)\\
    -\tau_{\textrm{c}} \chi_{1,2,1,1}(t)&\chi_{1,2,1,2}(t)&\cdots&-\tau_{\textrm{c}} \chi_{1,2,I,U_I}(t)\\
    \vdots & \vdots & \ddots & \vdots \\  
    -\tau_{\textrm{c}} \chi_{I,U_I,1,1}(t)&-\tau_{\textrm{c}} \chi_{I,U_I,1,2}(t)&\cdots&\chi_{I,U_I,I,U_I}(t)
    \end{bmatrix}}_{\triangleq \, \boldsymbol{\chi}(t)\in\mathbb{R}^{U\times U}}
    \underbrace{\begin{bmatrix}
    p_{1,1}^{\textrm{c}}(t) \\
    p_{1,2}^{\textrm{c}}(t) \\
    \vdots\\
    p_{I,U_I}^{\textrm{c}}(t)
    \end{bmatrix}}_{\triangleq \, \mathbf{p}(t)\in\mathbb{R}^{U}} = \underbrace{\begin{bmatrix}
      \tau_{\textrm{c}} \nu_{1,1}(t)
     \\
      \tau_{\textrm{c}} \nu_{1,2}(t) \\
      \vdots\\
       \tau_{\textrm{c}} \nu_{I,U_I}(t)
    \end{bmatrix}}_{\triangleq \, \boldsymbol{\nu}(t)\in\mathbb{R}^{U}}.
    \label{optimality condition} 
    \tag{23}
    \end{align}
    \hrulefill
\vspace{-4mm}
\end{figure*} 

\subsection{Sensing Beamforming Design}\label{Sec:Beamforming for Radar Sensing}
The objective of the sensing beamforming design is to enable $I$ satellites and $L$ gateways to probe all $M$ grid points over $T$ time slots. We first introduce a mapping function $\psi : \mathcal{T} \rightarrow \mathcal{M}$, under which the satellites and gateways are configured to probe the $\psi(t)$-th grid point at time slot $t$. Without careful beamforming design, the transmitted sensing and communication signals may interfere with each other. To address this issue, the sensing transmit and receive beamforming vectors for satellite $i$ and gateway $l$ at time slot $t$ are determined as
\begin{align}
\mathbf{f}_{i}^{\textrm{r}}(t)= \frac{1}{\sqrt{N^{\textrm{sat}}}} \mathbf{v}^{\textrm{sat-grid}}_{i,\psi(t)},~~\mathbf{w}_{l}(t)  = \frac{1}{\sqrt{N^{\textrm{gat}}}} \mathbf{v}^{\textrm{gat-grid}}_{l,\psi(t)},~\label{proposed radar transmit receive}
\end{align}
for $i\in\mathcal{I},\,l\in\mathcal{L},\,t\in\mathcal{T}$, respectively.

Following \cite{chen2013does}, we analyze the performance of the proposed beamforming design. Consider a half-wavelength spaced UPA with $N_x$ and $N_y$ elements along the $x$- and $y$-axes.
Let $\mathbf{v}(\phi,\theta)=\mathbf{v}_x(\phi,\theta)\otimes \mathbf{v}_y(\phi,\theta)\in\mathbb{C}^{N}$ with $N=N_xN_y$ denote the array steering vector toward azimuth $\phi$ and elevation $\theta$.
Then, for two directions $(\phi_1,\theta_1)$ and $(\phi_2,\theta_2)$, the crosstalk coefficient between their steering vectors is
\begin{align}
\frac{|\mathbf{v}(\phi_1,\theta_1)^H\mathbf{v}(\phi_2,\theta_2)|^2}{N}
\!=\! \frac{1}{N} \! \left|
\frac{\sin(\frac{\pi N_x\delta_x}{2})\sin(\frac{\pi N_y\delta_y}{2})}{\sin(\frac{\pi\delta_x}{2})\sin(\frac{\pi\delta_y}{2})}
\right|^2\!\!\!,
\end{align}
where $\delta_x = \cos\phi_1\cos\theta_1 - \cos\phi_2\cos\theta_2$ and $\delta_y = \sin\phi_1\cos\theta_1 - \sin\phi_2\cos\theta_2$.
Thus, if $(\phi_1,\theta_1)\neq(\phi_2,\theta_2)$ so that $(\delta_x,\delta_y)\neq(0,0)$, the crosstalk decays to zero as $N$ grows, implying asymptotic orthogonality of distinct steering vectors. As a result, when the array size is sufficiently large, the proposed beamforming designs in \eqref{proposed radar transmit receive} enable gateways to receive sensing signals without experiencing significant interference from communication signals. Motivated by this result, we allocate the maximum transmit power to each sensing beamforming vector, i.e., $p_{i}^{\textrm{r}}(t) = P_{i}^{\textrm{r}}$, $i\in\mathcal{I}$, $t\in\mathcal{T}$, where $P_{i}^{\textrm{r}}$ denotes the maximum sensing power of satellite $i$, in order to maximize the received sensing signal-to-noise ratio (SNR) at the gateways.
Note that our design also satisfies the constraints \eqref{hybrid architecture radar receive} and \eqref{hybrid architecture radar transmit} imposed by the hybrid beamforming architecture. 

\subsection{Communication Beamforming Design}\label{Sec:Beamforming for Communication UEs}
Following the same logic as in Sec. \ref{Sec:Beamforming for Radar Sensing}, the beamforming vector for the $u$-th UE served by satellite $i$ is given by
\begin{align}
\mathbf{f}_{i,u}^{\textrm{c}}(t)= \frac{1}{\sqrt{N^{\textrm{sat}}}}\mathbf{v}^{\textrm{sat-ue}}_{i,i,u},~~i\in\mathcal{I},\,u\in\mathcal{U}_i,\, t\in\mathcal{T}.
\end{align}
Conventional beamforming schemes such as weighted minimum mean square error \cite{christensen2008weighted} and regularized zero-forcing \cite{peel2005vector}, which are known to achieve optimal or near-optimal performance for sum-rate maximization under transmit power constraints, can be applied to downlink multi-user MIMO systems. However, they are typically designed for fully digital architectures and do not satisfy the hybrid beamforming constraints in \eqref{hybrid architecture communication transmit}. For this reason, we adopt a suboptimal beamforming strategy based on the LoS steering vector that satisfies the hybrid beamforming constraints and can mitigate sensing interference.

Next, we consider a power allocation problem to minimize the total power consumption under QoS constraints. The power minimization problem for time slot $t$ is formulated as
\begin{subequations}
\label{communication power minimization problem}
\begin{align}
\min_{\{p_{i,u}^{\textrm{c}}(t)\}_{i\in\mathcal{I},u\in\mathcal{U}_i}} ~~ &\sum_{i\in\mathcal{I}}\sum_{u\in\mathcal{U}_i}p_{i,u}^{\textrm{c}}(t)\\
\textrm{s.t.}~~~~~~~~~&\textrm{SINR}^{\textrm{ue}}_{i,u}(t)\geq \tau_{\textrm{c}},~~i\in\mathcal{I},\,u\in\mathcal{U}_i,\label{SINR constraints}\\
&p_{i,u}^{\textrm{c}}(t) \geq 0,~~i\in\mathcal{I},\,u\in\mathcal{U}_i,\label{non-negative power constraints}
\end{align}
\end{subequations}
with $\tau_{\textrm{c}}$ representing the SINR threshold for communication. For solving \eqref{communication power minimization problem}, we first rewrite the problem by defining $\chi_{i,u,j,v}(t)= \left|\mathbf{h}^H_{j,i,u}(t)\mathbf{f}_{j,v}^{\textrm{c}}(t)\right|^2$ and $\nu_{i,u}(t) = \sum\limits_{j\in\mathcal{I}}P_{j}^{\textrm{r}}\left|\mathbf{h}^H_{j,i,u}(t)\mathbf{f}_{j}^{\textrm{r}}(t)\right|^2+\sigma_{i,u}^2$. 
Applying this, constraint \eqref{SINR constraints} can be transformed into the following equivalent constraints:
\begin{align}
p_{i,u}^{\textrm{c}}(t)\chi_{i,u,i,u}(t) \geq& \sum\limits_{j\in\mathcal{I}\setminus\left\{i\right\}}\sum\limits_{v\in\mathcal{U}_j}\tau_{\textrm{c}} p_{j,v}^{\textrm{c}}(t)\chi_{i,u,j,v}(t) + \tau_{\textrm{c}}\nu_{i,u}(t) \nonumber\\
&+ \sum\limits_{v\in\mathcal{U}_i\setminus\left\{u\right\}}\tau_{\textrm{c}}p_{i,v}^{\textrm{c}}(t)\chi_{i,u,i,v}(t),\label{SINR constraints LP}
\end{align}
for all $i\in\mathcal{I},\,u\in\mathcal{U}_i$. Thus, the problem is reformulated as
\begin{subequations}
\label{communication power minimization problem LP}
\begin{align}
\min_{\{p_{i,u}^{\textrm{c}}(t)\}_{i\in\mathcal{I},u\in\mathcal{U}_i}} ~~ &\sum_{i\in\mathcal{I}}\sum_{u\in\mathcal{U}_i}p_{i,u}^{\textrm{c}}(t)\\
\textrm{s.t.} ~~~~~~~~~&\eqref{non-negative power constraints},\, \eqref{SINR constraints LP},
\end{align}
\end{subequations}
which is a standard linear programming (LP) that can be efficiently solved \cite{jo2026beam}.

A feasible and bounded LP always has an optimal solution at a vertex of the feasible polyhedron \cite{boyd2004convex}, which implies that the optimal solution to the proposed problem can be achieved when \eqref{SINR constraints LP} holds with equality. As a result, the optimality condition for problem \eqref{communication power minimization problem LP} can be derived as shown in \eqref{optimality condition} at the top of this page, where $U = \sum_{i\in\mathcal{I}}U_i$.
According to the L\'evy-Desplanques theorem \cite{horn2012matrix}, any strictly diagonally dominant square matrix is non-singular, which means that the matrix $\boldsymbol{\chi}(t)$ is invertible when it satisfies the following conditions:
\setcounter{equation}{23}
\begin{align}
\chi_{i,u,i,u}(t) & > \sum\limits_{j\in\mathcal{I}\setminus\left\{i\right\}}\sum\limits_{v\in\mathcal{U}_j}\tau_{\textrm{c}} \chi_{i,u,j,v}(t) + \sum\limits_{v\in\mathcal{U}_i\setminus\left\{u\right\}}\tau_{\textrm{c}}\chi_{i,u,i,v}(t),
\label{Communication power opt condition}
\end{align}
for all $i\in\mathcal{I},\,u\in\mathcal{U}_i$.
However, the inequalities in \eqref{Communication power opt condition} may not always hold, particularly when $\tau_{\textrm{c}}$ is chosen too large. A practical approach is to adjust the communication SINR threshold to a feasible range. Under this adjustment, the feasibility of problem \eqref{communication power minimization problem LP} is guaranteed, and the optimal solution is given by $\mathbf{p}(t)=\boldsymbol{\chi}^{-1}(t)\boldsymbol{\nu}(t)$.

\section{Proposed Centralized Multi-Target Sensing Framework}\label{Sec:Centralized Sensing Framework}
In this section, we present a centralized sensing framework for multi-satellite ISAC networks, where the observations from all gateways are collected and jointly processed to estimate target positions in a 3D region. To this end, we formulate a centralized target sensing problem to exploit the collected observations. \textcolor{black}{We then introduce two target sensing algorithms for this problem: grid-based and grid-free algorithms.}

\subsection{Centralized Sensing Problem}\label{Sec:Target Sensing Problem}
In practice, centralized processing can be realized by employing a cloud radio access network architecture \cite{simeone2016cloud}. In this setup, the observations from distributed gateways are transmitted to the CU and jointly processed. We define $\mathbf{y}_{l}\in\mathbb{C}^{T}$ as a sensing observation vector that concatenates the observations at gateway $l$ over $T$ time slots as shown in \eqref{measurement model} at the top of page 7.
\begin{figure*}[!ht]
\vspace{-5mm}
    \begin{align}
    \underbrace{\begin{bmatrix}
      y_{l}^{\textrm{gat}}(1) \\
      y_{l}^{\textrm{gat}}(2) \\
      \vdots\\
      y_{l}^{\textrm{gat}}(T)
    \end{bmatrix}}_{\triangleq \, \mathbf{y}_{l}\in\mathbb{C}^{T}} 
    =&\underbrace{
    \begin{bmatrix}
    a_{1,1,l}^{\textrm{gat}}(1)&\cdots&a_{1,K,l}^{\textrm{gat}}(1)&\cdots&a_{I,1,l}^{\textrm{gat}}(1)&\cdots&a_{I,K,l}^{\textrm{gat}}(1)\\
    a_{1,1,l}^{\textrm{gat}}(2)&\cdots&a_{1,K,l}^{\textrm{gat}}(2)&\cdots&a_{I,1,l}^{\textrm{gat}}(2)&\cdots&a_{I,K,l}^{\textrm{gat}}(2)\\
    \vdots &\ddots& \vdots & \ddots & \vdots & \ddots & \vdots\\  
    a_{1,1,l}^{\textrm{gat}}(T)&\cdots&a_{1,K,l}^{\textrm{gat}}(T)&\cdots&a_{I,1,l}^{\textrm{gat}}(T)&\cdots&a_{I,K,l}^{\textrm{gat}}(T)
    \end{bmatrix}}_{\triangleq \, \mathbf{A}^{\textrm{gat}}_{l}\in\mathbb{C}^{T\times KI}}  \underbrace{\begin{bmatrix}
    \rho_{1,1,l} \\
    \vdots\\
    \rho_{1,K,l} \\
    \vdots\\
    \rho_{I,1,l} \\
    \vdots\\
    \rho_{I,K,l} 
    \end{bmatrix}}_{\triangleq \, \boldsymbol{\rho}_l\in\mathbb{C}^{KI}} + \underbrace{\begin{bmatrix}
       \mathbf{w}^{H}_{l}(1)\mathbf{n}^{\textrm{gat}}_l(1)
     \\
       \mathbf{w}^{H}_{l}(2)\mathbf{n}^{\textrm{gat}}_l(2) \\
      \vdots\\
       \mathbf{w}^{H}_{l}(T)\mathbf{n}^{\textrm{gat}}_l(T)
    \end{bmatrix}}_{\triangleq \, \boldsymbol{\omega}_{l}\in\mathbb{C}^{T}}.
    \label{measurement model} 
    \end{align}
    \hrulefill
\vspace{-4mm}
\end{figure*} 
The observation vector $\mathbf{y}_{l}$ is forwarded to the CU and combined with those from other gateways for joint processing. Stacking the observations of all gateways yields
\begin{align}
\underbrace{\begin{bmatrix}
  \mathbf{y}_{1} \\
  \mathbf{y}_{2} \\
  \vdots\\
  \mathbf{y}_{L}
\end{bmatrix}}_{\triangleq \, \mathbf{y}\in\mathbb{C}^{TL}} \!\!=\!
\underbrace{
\begin{bmatrix}
  \mathbf{A}_1^{\textrm{gat}} & 0 & \cdots & 0 \\
  0 & \mathbf{A}_2^{\textrm{gat}} & \cdots & 0 \\
  \vdots & \vdots & \ddots & \vdots \\
  0 & 0 & \cdots & \mathbf{A}_L^{\textrm{gat}}
\end{bmatrix}}_{\triangleq \, \mathbf{A}^{\textrm{gat}}\in\mathbb{C}^{TL\times KIL}}
\!\!\!\underbrace{\begin{bmatrix}
\boldsymbol{\rho}_1\\
\boldsymbol{\rho}_2\\
\vdots\\
\boldsymbol{\rho}_L\\
\end{bmatrix}}_{\triangleq \, \boldsymbol{\rho}\in\mathbb{C}^{KIL}} \!\!+\!\! \underbrace{\begin{bmatrix}
  \boldsymbol{\omega}_{1}
 \\
 \boldsymbol{\omega}_{2} \\
  \vdots\\
 \boldsymbol{\omega}_{L}
\end{bmatrix}}_{\triangleq \, \boldsymbol{\omega}\in\mathbb{C}^{TL}}.
\label{centralized received signal}
\end{align}
Based on this stacked observation $\mathbf{y}$, the CU estimates the $K$ unknown target positions $\{\mathbf{q}^{\textrm{tar}}_{k}\}_{k\in\mathcal{K}}$. This estimation task can be formulated as a sparse signal recovery problem since only a limited number of targets exist within the search space. 

\subsection{Grid-Based Sensing Algorithm}\label{Sec:Grid-Based Sensing Algorithm}
Following the grid structure described in Sec. \ref{Sec:Grid Structure for Target Sensing}, we introduce the grid-based sensing matrix $\mathbf{A}^{\textrm{grid}}\in\mathbb{C}^{TL\times MIL}$ as 
\begin{align}
\mathbf{A}^{\textrm{grid}}=\textrm{blkdiag}(\mathbf{A}_1^{\textrm{grid}},\cdots,\mathbf{A}_L^{\textrm{grid}}),
\end{align}
where $\mathbf{A}_l^{\textrm{grid}}\in\mathbb{C}^{T\times MI}$ is defined as
\begin{align}
{\setlength{\arraycolsep}{1.5pt}
\mathbf{A}_l^{\textrm{grid}} \!=\! \begin{bmatrix}
a_{1,1,l}^{\textrm{grid}}(1)&\cdots&a_{1,M,l}^{\textrm{grid}}(1)&\cdots&a_{I,1,l}^{\textrm{grid}}(1)&\cdots&a_{I,M,l}^{\textrm{grid}}(1)\\
a_{1,1,l}^{\textrm{grid}}(2)&\cdots&a_{1,M,l}^{\textrm{grid}}(2)&\cdots&a_{I,1,l}^{\textrm{grid}}(2)&\cdots&a_{I,M,l}^{\textrm{grid}}(2)\\
\vdots &\ddots& \vdots & \ddots & \vdots & \ddots & \vdots\\  
a_{1,1,l}^{\textrm{grid}}(T)&\cdots&a_{1,M,l}^{\textrm{grid}}(T)&\cdots&a_{I,1,l}^{\textrm{grid}}(T)&\cdots&a_{I,M,l}^{\textrm{grid}}(T)\\
\end{bmatrix}.}
\end{align}
With the grid-based sensing matrix $\mathbf{A}^{\textrm{grid}}$, the stacked observation $\mathbf{y}$ can be approximated as
\begin{align}
\mathbf{y} \approx \mathbf{A}^{\textrm{grid}}
\boldsymbol{\rho}^{\textrm{grid}} + \boldsymbol{\omega},
\label{grid-based approximate model}
\end{align}
where $\boldsymbol{\rho}^{\textrm{grid}}\in\mathbb{C}^{MIL}$ is a sparse vector with $KIL$ non-zero entries. Specifically, if a target is located on the $m$-th grid point, the corresponding entries are non-zero, i.e., $[\boldsymbol{\rho}^{\textrm{grid}}]_{m+rM}\neq0$ for each $r \in \mathcal{R} \triangleq \{0,1,\cdots,IL-1\}$. 
Considering this, we formulate the grid-based sensing problem that estimates the sparse signal $\boldsymbol{\rho}^{\textrm{grid}}$ from its noisy observation $\mathbf{y}$ as  
\begin{subequations}
\label{cooperative sensing problem}
\begin{align}
\min_{\boldsymbol{\rho}^{\textrm{grid}}\in\mathbb{C}^{MIL}}  ~~  &\big\|\mathbf{y}-\mathbf{A}^{\textrm{grid}}\boldsymbol{\rho}^{\textrm{grid}}\big\|_2^2  \\
\text{s.t.}~~~~~\,\,&  \sum_{m=1}^M\mathbb{I}\big[\|\boldsymbol{\rho}^{\textrm{group}}_{m}\|_0 = IL\big] = K, \label{cooperative_constraint}
\end{align}
\end{subequations}
where $\boldsymbol{\rho}^{\textrm{group}}_{m}= \big[[\boldsymbol{\rho}^{\textrm{grid}}]_{m},[\boldsymbol{\rho}^{\textrm{grid}}]_{m+M},\cdots,[\boldsymbol{\rho}^{\textrm{grid}}]_{m+(IL-1)M}\big]$ and $\mathbb{I}[\cdot]$ is the indicator function. 

\textcolor{black}{The approximation in \eqref{grid-based approximate model} stems from two factors. First, the grid-based sensing matrix neglects the reflected components induced by communication signals. This is justified by the asymptotic orthogonality between distinct steering directions. Second,} in practical environments, it is very unlikely that targets are located exactly on grid points. This mismatch error can be reduced by increasing the grid resolution within the sensing region. However, a larger number of grid points not only increases the complexity of signal recovery, but also reduces its performance by weakening the restricted isometry property \cite{nguyen2025performance, eldar2012compressed}. Consequently, the number of grid points should be selected to achieve a desirable trade-off between grid mismatch error and recovery performance degradation.


The grid-based target sensing problem in \eqref{cooperative sensing problem} is NP-hard, as it requires a combinatorial search over all possible sparsity patterns under the constraint in \eqref{cooperative_constraint}. To avoid the prohibitive complexity of exhaustive search, several efficient algorithms have been developed in the literature \cite{shi2022device}, \cite{yang2025cooperative}, \cite{shi2024joint}, \cite{jo2026beam}. Among these algorithms, we adopt the well-known orthogonal matching pursuit (OMP) algorithm as it offers a significant advantage in terms of computational complexity. It should be noted that more advanced sparse recovery algorithms beyond OMP can also be employed to solve \eqref{cooperative sensing problem}; however, such methods generally incur higher computational complexity. Moreover, the numerical results in Sec. \ref{Sec:Numerical Results} demonstrate that OMP is sufficient to illustrate the superiority of the proposed sensing framework over the considered baseline schemes.

{\setlength{\textfloatsep}{1pt}
\begin{algorithm}[t] 
    \caption{OMP-Based Centralized Sensing Algorithm}\label{alg:omp-cent-sensing}
    {
    \small
    {\begin{algorithmic}[1]
        \REQUIRE $\mathbf{y}$, $\mathbf{A}^{\textrm{grid}}$, $K$.
        \ENSURE  $\hat{\mathcal{M}}$.
        \STATE Initialize $\boldsymbol{\Delta}^{(0)}=\mathbf{y}$ and $\hat{\mathcal{M}}^{(0)}=\varnothing$.
        \FOR{$k=1:K$}
        \STATE $\hat{m}_k = \underset{m \in \mathcal{M}\setminus \hat{\mathcal{M}}^{(k-1)}}{\arg\!\max} ~\underset{r \in \mathcal{R}}{\sum}\left|\big(\boldsymbol{\Delta}^{(k-1)}\big)^H[\mathbf{A}^{\textrm{grid}}]_{:,m+rM}\right|$.
        \STATE $\hat{\mathcal{M}}^{(k)} = \hat{\mathcal{M}}^{(k-1)} \cup \{\hat{m}_k\}$.
        \STATE $\mathbf{A}^{(k)} = \big[[\mathbf{A}^{\textrm{grid}}]_{:,m+rM}\big]_{m\in\hat{\mathcal{M}}^{(k)}, r \in \mathcal{R}}$.
        \STATE $\boldsymbol{\Delta}^{(k)} = \mathbf{y}-\mathbf{A}^{(k)}(\mathbf{A}^{(k)})^{\dagger}\mathbf{y}$.
        \ENDFOR
        \STATE \textbf{Return} $\hat{\mathcal{M}} = \hat{\mathcal{M}}^{(K)}$.
    \end{algorithmic}}
    }
\end{algorithm}
}

The fundamental of the OMP algorithm is to iteratively select the column of the measurement matrix that is most correlated with the current residual and update the estimate accordingly. First, we initialize the residual vector as $\boldsymbol{\Delta}^{(0)}=\mathbf{y}$ and the solution set as $\hat{\mathcal{M}}^{(0)}=\varnothing$.
At the $k$-th iteration, the grid point $\hat{m}_k$ most correlated with the current residual is selected as
\begin{align}
\hat{m}_k = \underset{m \in \mathcal{M}\setminus \hat{\mathcal{M}}^{(k-1)}}{\arg\!\max}~ \underset{r \in \mathcal{R}}{\sum}\left|\big(\boldsymbol{\Delta}^{(k-1)}\big)^H[\mathbf{A}^{\textrm{grid}}]_{:,m+rM}\right|.
\end{align}
\textcolor{black}{Note that the $\ell_2$-norm can be dominated by a single large correlation, which fails to fully exploit the advantages of the cooperative framework. We thus adopt the $\ell_1$-norm to capture the correlation without overemphasizing any component.}
Then, the solution is updated based on the selected grid point, as $\hat{\mathcal{M}}^{(k)} = \hat{\mathcal{M}}^{(k-1)} \cup \{\hat{m}_k\}$. Subsequently, the residual is updated by subtracting from $\mathbf{y}$ its projection onto the subspace spanned by the selected columns, which is given by
\begin{align}
\boldsymbol{\Delta}^{(k)} = \mathbf{y}-\mathbf{A}^{(k)}(\mathbf{A}^{(k)})^{\dagger}\mathbf{y},
\end{align}
where $\mathbf{A}^{(k)} = [[\mathbf{A}^{\textrm{grid}}]_{:,m+rM}]_{m\in\hat{\mathcal{M}}^{(k)}, r \in \mathcal{R}}$ and $(\mathbf{A}^{(k)})^{\dagger}=( (\mathbf{A}^{(k)})^H\mathbf{A}^{(k)})^{-1}(\mathbf{A}^{(k)})^H$. This greedy procedure repeats until a predefined stopping criterion is satisfied. Given prior knowledge of the number of targets, the convergence condition is $|\hat{\mathcal{M}}^{(k)}|=K$; otherwise, it can be stopped when $\|\boldsymbol{\Delta}^{(k)}\|_2 \leq \epsilon$ for a given threshold $\epsilon$. In this study, perfect prior knowledge is assumed to focus the analysis on sensing performance without loss due to target number uncertainty. The details of the proposed algorithm are summarized in Algorithm \ref{alg:omp-cent-sensing}. After executing the above procedure, the estimated positions of the multiple targets are determined as
\begin{align}
\{\hat{\mathbf{q}}^{\textrm{tar}}_{k}\}_{k\in\mathcal{K}} = \{\mathbf{q}^{\textrm{grid}}_{\hat{m}_k}\}_{\hat{m}_k \in \hat{\mathcal{M}}}.
\end{align}

{\color{black}
\subsection{Grid-Free Sensing Algorithm}\label{Sec:Grid-Free Sensing Algorithm}
}
\textcolor{black}{
The grid-based sensing algorithm provides an efficient framework for sparse recovery. However, its sensing accuracy is inevitably limited by the predefined grid structure. To address this limitation, we develop a grid-free sensing algorithm that directly searches for target positions in the continuous sensing region.}

\textcolor{black}{For an arbitrary position $\mathbf{q}\in\mathbb{R}^3$, we define the continuous sensing matrix $\mathbf{A}^{\textrm{cont}}(\mathbf{q})\in\mathbb{C}^{TL\times IL}$ as
\begin{align}
\mathbf{A}^{\textrm{cont}}(\mathbf{q})=\textrm{blkdiag}(\mathbf{A}_1^{\textrm{cont}}(\mathbf{q}),\cdots,\mathbf{A}_L^{\textrm{cont}}(\mathbf{q})).
\end{align}
In this equation, $\mathbf{A}_l^{\textrm{cont}}(\mathbf{q}) \in \mathbb{C}^{T\times I}$ denotes the submatrix for gateway $l$, whose $(t,i)$-th entry is given by
\begin{align}
[\mathbf{A}_l^{\textrm{cont}}(\mathbf{q})]_{t,i} =&\sqrt{\frac{\lambda^2 A^{\textrm{gat}}_{\textrm{Rx}}A^{\textrm{sat}}_{\textrm{Tx}}p_{i}^{\textrm{r}}(t)}{64\pi^3 (d^{\textrm{gat}}_{l}(\mathbf{q}))^2(d^{\textrm{sat}}_{i}(\mathbf{q}))^2}}e^{-j2\pi\frac{ d^{\textrm{gat}}_{l}(\mathbf{q})+d^{\textrm{sat}}_{i}(\mathbf{q})}{\lambda}}\nonumber\\
&\times \mathbf{w}^{H}_{l}(t)\mathbf{v}^{\textrm{gat}}_{l}(\mathbf{q}) (\mathbf{v}^{\textrm{sat}}_{i}(\mathbf{q}))^{H}  \mathbf{f}_{i}^{\textrm{r}}(t)s_{i}^{\textrm{r}}(t),
\end{align}
where $d^{\textrm{gat}}_{l}(\mathbf{q}) = \| \mathbf{q}^{\textrm{gat}}_{l}-\mathbf{q}\|_{2}$ and $d^{\textrm{sat}}_{i}(\mathbf{q}) = \| \mathbf{q}^{\textrm{sat}}_{i}-\mathbf{q}\|_{2}$ are the distances from position $\mathbf{q}$ to gateway $l$ and satellite $i$, respectively. The array response vectors for these links are denoted by $\mathbf{v}^{\textrm{gat}}_{l}(\mathbf{q})$ and $\mathbf{v}^{\textrm{sat}}_{i}(\mathbf{q})$, respectively. From the definition of $\mathbf{A}^{\textrm{cont}}(\mathbf{q})$, the stacked observation $\mathbf{y}$ in \eqref{centralized received signal} can be rewritten as
\begin{align}
\mathbf{y} = \sum_{k\in\mathcal{K}}\mathbf{A}^{\textrm{cont}}(\mathbf{q}_k^{\textrm{tar}}) \boldsymbol{\rho}^{\textrm{cont}}_k + \boldsymbol{\omega},
\end{align}
with $\boldsymbol{\rho}^{\textrm{cont}}_k = [\rho_{1,k,1},\cdots,\rho_{I,k,1},\cdots,\rho_{1,k,L},\cdots,\rho_{I,k,L}]^T$.
Given the noisy observation $\mathbf{y}$, the target positions are estimated by solving the following grid-free sensing problem:
\begin{subequations}
\begin{align}
\min_{\{\mathbf{q}_k\}_{k\in\mathcal{K}},\{\boldsymbol{\rho}^{\textrm{cont}}_k\}_{k\in\mathcal{K}}} ~~ &\big\|\mathbf{y}-\sum_{k\in\mathcal{K}}\mathbf{A}^{\textrm{cont}}(\mathbf{q}_k) \boldsymbol{\rho}^{\textrm{cont}}_k\big\|_2^2 \\
\textrm{s.t.} ~~~~~~~~~~&\mathbf{q}_k \in \mathbb{R}^3,\,\boldsymbol{\rho}^{\textrm{cont}}_k\in\mathbb{C}^{IL},~~k\in\mathcal{K}.
\end{align}
\end{subequations}
}

{\setlength{\textfloatsep}{1pt}
\begin{algorithm}[t] 
\color{black}
    \caption{\textcolor{black}{PSO-Based Centralized Sensing Algorithm}}\label{alg:pso-cent-sensing}
    {
    \small
    {\begin{algorithmic}[1]
        \REQUIRE $\mathbf{y}$, $\{\mathbf{q}^{\textrm{sat}}_{i}\}_{i\in\mathcal{I}}$, $\{\mathbf{q}^{\textrm{gat}}_{l}\}_{l\in\mathcal{L}}$, $K$, $P$, $N$, $w_{\max}$, $w_{\min}$, $c_1$, $c_2$.
        \ENSURE  $\{\hat{\mathbf{q}}^{\textrm{tar}}_{k}\}_{k\in\mathcal{K}}$.
        \STATE Initialize $\boldsymbol{\Delta}^{(0)}=\mathbf{y}$.
        \FOR{$k=1:K$}
        \STATE Initialize particles $\{\mathbf{p}_{p}^{(k,0)}\}_{p\in\{1,\cdots,P\}}$ in the sensing region.
        \STATE Set $\mathbf{v}_{p}^{(k,0)}=\mathbf{0}_{3}$ and $\hat{\mathbf{p}}_{p}^{(k)}=\mathbf{p}_{p}^{(k,0)}$ for all $p\in\{1,\cdots,P\}$.
        \STATE Find $\hat{\mathbf{g}}^{(k)}\!=\arg\!\max_{\{\mathbf{p}_{p}^{(k,0)}\}_{p\in\{1,\cdots,P\}}} F(\boldsymbol{\Delta}^{(k-1)},\mathbf{p}_{p}^{(k,0)})$.
            \FOR{$n=1:N$}
            \STATE $w^{(k,n)}=w_{\max}-\frac{w_{\max}-w_{\min}}{N-1}(n-1)$.
                \FOR{$p=1:P$}
                \STATE Randomly generate $\mathbf{r}_1^{(k,n)},\mathbf{r}_2^{(k,n)} \in [0,1]^3$.
                \STATE $\mathbf{v}_{p}^{(k,n)} = w^{(k,n)}\mathbf{v}_{p}^{(k,n-1)} \!+\! c_1 \mathbf{r}_1^{(k,n)} \odot(\hat{\mathbf{p}}_{p}^{(k)}-\mathbf{p}_{p}^{(k,n-1)})$ 
                \Statex $~~~~~~~~~~~~~~~~~+ c_2\mathbf{r}_2^{(k,n)}\odot(\hat{\mathbf{g}}^{(k)}-\mathbf{p}_{p}^{(k,n-1)}).$ 
                \STATE $\mathbf{p}_{p}^{(k,n)} = \mathbf{p}_{p}^{(k,n-1)} +\mathbf{v}_{p}^{(k,n)}$.
                \STATE Adjust $\mathbf{p}_{p}^{(k,n)}$ if it is outside the sensing region.
                \IF{$F(\boldsymbol{\Delta}^{(k-1)},\mathbf{p}_{p}^{(k,n)}) > F(\boldsymbol{\Delta}^{(k-1)},\hat{\mathbf{p}}_{p}^{(k)})$}
                \STATE $\hat{\mathbf{p}}_{p}^{(k)}=\mathbf{p}_{p}^{(k,n)}$.
                \ENDIF
                \IF{$F(\boldsymbol{\Delta}^{(k-1)},\mathbf{p}_{p}^{(k,n)}) > F(\boldsymbol{\Delta}^{(k-1)},\hat{\mathbf{g}}^{(k)})$}
                \STATE $\hat{\mathbf{g}}^{(k)}=\mathbf{p}_{p}^{(k,n)}$.
                \ENDIF
                \ENDFOR
            \ENDFOR
            \STATE $\hat{\mathbf{q}}^{\textrm{tar}}_{k} = \hat{\mathbf{g}}^{(k)}$.
            \STATE $\mathbf{A}^{(k)} = \big[\mathbf{A}^{\textrm{cont}}(\hat{\mathbf{q}}^{\textrm{tar}}_{1}),\cdots,\mathbf{A}^{\textrm{cont}}(\hat{\mathbf{q}}^{\textrm{tar}}_{k})\big]$.
            \STATE $\boldsymbol{\Delta}^{(k)} = \mathbf{y}-\mathbf{A}^{(k)}(\mathbf{A}^{(k)})^{\dagger}\mathbf{y}$.
        \ENDFOR
        \STATE \textbf{Return} $\{\hat{\mathbf{q}}^{\textrm{tar}}_{k}\}_{k\in\mathcal{K}}$.
    \end{algorithmic}}
    }
\end{algorithm}
}

\vspace{-5mm}
\textcolor{black}{
This grid-free sensing problem is intractable due to its non-convex objective function.
We adopt the PSO algorithm \cite{wang2025multiple} to handle the continuous non-convex search space.  
Specifically, it is a stochastic optimization method that mimics the collective intelligence of biological swarms.
The search space is iteratively explored by a group of particles whose positions represent potential solutions.
During the iterative search process, each particle evaluates the fitness value at its current position and updates its position based on both its own best position and the best position found by the entire swarm. This collective search behavior facilitates efficient exploration with improved robustness against local optima and fast convergence. The PSO algorithm is also suitable for complex optimization problems, as it has a simple mathematical structure with only a few tuning parameters. Motivated by these advantages, we develop a PSO-based grid-free sensing algorithm that sequentially estimates target positions by updating the residual signal.
}

\textcolor{black}{For sequential target estimation, as in Sec. \ref{Sec:Grid-Based Sensing Algorithm}, we first define the residual vector. Suppose that the first $k-1$ target positions have already been estimated as $\hat{\mathbf{q}}^{\textrm{tar}}_{1},\cdots,\hat{\mathbf{q}}^{\textrm{tar}}_{k-1}$. Then, the continuous sensing matrix constructed from the estimated target positions is given by
\begin{align}
\mathbf{A}^{(k-1)} = \big[\mathbf{A}^{\textrm{cont}}(\hat{\mathbf{q}}^{\textrm{tar}}_{1}),\cdots,\mathbf{A}^{\textrm{cont}}(\hat{\mathbf{q}}^{\textrm{tar}}_{k-1})\big].
\end{align}
Given the matrix $\mathbf{A}^{(k-1)}$, the residual signal after removing the previously detected target components via the least-squares projection is obtained as
\begin{align}
\boldsymbol{\Delta}^{(k-1)} = \mathbf{y}-\mathbf{A}^{(k-1)}(\mathbf{A}^{(k-1)})^{\dagger}\mathbf{y}.
\end{align}
Based on the residual $\boldsymbol{\Delta}^{(k-1)}$, we apply the PSO algorithm to
estimate the $k$-th target position. Let $\mathbf{p}_{p}^{(k,n)}\in\mathbb{R}^{3}$ be the position of the $p$-th particle at the $n$-th PSO iteration. The fitness of this particle is evaluated as
\begin{align}
F(\boldsymbol{\Delta}^{(k-1)},\mathbf{p}_{p}^{(k,n)}) =&-\Big\|\boldsymbol{\Delta}^{(k-1)}-\mathbf{A}^{\textrm{cont}}(\mathbf{p}_{p}^{(k,n)})\nonumber\\
&\times(\mathbf{A}^{\textrm{cont}}(\mathbf{p}_{p}^{(k,n)}))^{\dagger}\boldsymbol{\Delta}^{(k-1)}
\Big\|_2^2.
\end{align}
This fitness function measures how well the particle position fits the current residual signal. At each iteration, the personal best position of each particle and the global best position of the swarm are updated according to the fitness values. Let $\hat{\mathbf{p}}_{p}^{(k)}$ and $\hat{\mathbf{g}}^{(k)}$ denote the personal best position of the $p$-th particle and the global best position, respectively. Then, the velocity and position of the $p$-th particle are updated as
\begin{align}
\mathbf{v}_{p}^{(k,n)} &= w^{(k,n)}\mathbf{v}_{p}^{(k,n-1)} + c_1 \mathbf{r}_1^{(k,n)} \odot(\hat{\mathbf{p}}_{p}^{(k)}-\mathbf{p}_{p}^{(k,n-1)}) \nonumber\\
&~~~+ c_2\mathbf{r}_2^{(k,n)}\odot(\hat{\mathbf{g}}^{(k)}-\mathbf{p}_{p}^{(k,n-1)}),\\
\mathbf{p}_{p}^{(k,n)} &= \mathbf{p}_{p}^{(k,n-1)} +\mathbf{v}_{p}^{(k,n)},
\end{align}
where $w^{(k,n)}$ is the inertia weight that linearly decreases from $w_{\max}$ to $w_{\min}$ over the PSO iterations, $c_1$ and $c_2$ are the acceleration coefficients, $\mathbf{r}_1^{(k,n)},\mathbf{r}_2^{(k,n)}\in[0,1]^3$ are random vectors, and $\odot$ denotes the element-wise product. After $N$ iterations, the global best position $\hat{\mathbf{g}}^{(k)}$ is selected as the estimate of the $k$-th target position. This procedure is repeated until the $K$ target positions are estimated. The detailed algorithm is summarized in Algorithm \ref{alg:pso-cent-sensing}, where $P$ denotes the number of particles.
}

\textcolor{black}{
\subsection{CRLB Analysis}\label{Sec:Theoretical Analysis}
We provide a theoretical analysis by deriving the CRLB for target sensing. This provides a lower bound on the estimation error covariance matrix of any unbiased estimator for the target locations.
In this analysis, the reflection coefficients are considered as deterministic unknown parameters.
We define the unknown parameters as $\boldsymbol{\eta}=[\boldsymbol{\vartheta}^T,\boldsymbol{\rho}_\textrm{R}^T,\boldsymbol{\rho}_\textrm{I}^T]^T \in \mathbb{R}^{3K+2KIL}$ with $\boldsymbol{\vartheta}=[(\mathbf{q}^{\textrm{tar}}_{1})^T,\cdots,(\mathbf{q}^{\textrm{tar}}_{K})^T]^T$, $\boldsymbol{\rho}_\textrm{R}=\textrm{Re}(\boldsymbol{\rho})$, and $\boldsymbol{\rho}_\textrm{I}=\textrm{Im}(\boldsymbol{\rho})$. 
Following \cite{kay1993statistical}, the Fisher information matrix (FIM) for estimating $\boldsymbol{\eta}$ from \eqref{centralized received signal} is given by
\begin{align}
\mathbf{J}_{\boldsymbol{\eta}}=
\begin{bmatrix}
\mathbf{J}_{\boldsymbol{\vartheta}\boldsymbol{\vartheta}} & \mathbf{J}_{\boldsymbol{\vartheta}\boldsymbol{\xi}}\\
\mathbf{J}_{\boldsymbol{\vartheta}\boldsymbol{\xi}}^T & \mathbf{J}_{\boldsymbol{\xi}\boldsymbol{\xi}} 
\end{bmatrix}, \label{FIM matrix}
\end{align}
where the $(m,n)$-th entry is defined as
\begin{align}
[\mathbf{J}_{\boldsymbol{\eta}}]_{m,n} = 2\sum_{l\in\mathcal{L}}\sum_{t\in\mathcal{T}}\frac{1}{\zeta_l^2}\textrm{Re}\left(\frac{\partial \mu_{l}(t)^H}{\partial[\boldsymbol{\eta}]_m}\frac{\partial \mu_{l}(t)}{\partial[\boldsymbol{\eta}]_n}\right)\label{FIM}
\end{align}
with $\boldsymbol{\xi}= [\boldsymbol{\rho}_\textrm{R}^T,\boldsymbol{\rho}_\textrm{I}^T]^T$ and $\mu_{l}(t)=\sum_{k\in\mathcal{K}}\sum_{i\in\mathcal{I}} \rho_{i,k,l}  a_{i,k,l}^{\textrm{gat}}(t)$.
The detailed derivation of the FIM is provided in Appendix A. Applying the Schur complement \cite{bekkerman2006target}, the equivalent FIM for $\boldsymbol{\vartheta}$ is obtained as $\mathbf{J}_{\boldsymbol{\vartheta}} = \mathbf{J}_{\boldsymbol{\vartheta}\boldsymbol{\vartheta}} - \mathbf{J}_{\boldsymbol{\vartheta}\boldsymbol{\xi}}\mathbf{J}_{\boldsymbol{\xi}\boldsymbol{\xi}}^{-1}\mathbf{J}_{\boldsymbol{\vartheta}\boldsymbol{\xi}}^T$. 
Thus, the CRLB for the target position vector $\boldsymbol{\vartheta}$ is given by 
\begin{align}\label{eq:CRLB}
\textrm{CRLB}(\boldsymbol{\vartheta}) = \mathbf{J}_{\boldsymbol{\vartheta}}^{-1}.
\end{align}
It is worthwhile to mention that the CRLB is determined by the Fisher information associated with the observation model rather than by the sensing algorithm itself.}


\section{Proposed Distributed Multi-Target Sensing Framework}\label{Sec:Distributed Sensing Framework}
The centralized framework jointly processes observations from all gateways to improve sensing performance. However, this entails substantial fronthaul overhead and high computational complexity. To overcome these limitations, we propose a distributed sensing framework that avoids transmitting raw observations by allowing each gateway to forward only its local sensing results to the CU.

\subsection{Non-Cooperative Sensing at Individual Gateway}\label{Sec:Local Target Sensing}
As an ingredient for developing the distributed sensing framework, we start by introducing a non-cooperative sensing framework in which target locations are individually estimated by each gateway from its local observations. 
The non-cooperative grid-based sensing problem for gateway $l$ is then formulated as 
\begin{subequations}
\label{Local Grid Target Sensing}
\begin{align}
\min_{\boldsymbol{\rho}^{\textrm{grid}}_l \in \mathbb{C}^{MI} }  ~~  &\big\|\mathbf{y}_l-\mathbf{A}^{\textrm{grid}}_l\boldsymbol{\rho}^{\textrm{grid}}_l\big\|_2^2 \\
\text{s.t.}~~~~~&  \sum_{m=1}^M\mathbb{I}\big[\big\|\boldsymbol{\rho}^{\textrm{group}}_{l,m}\big\|_0= I\big] = K,
\end{align}
\end{subequations}
where $\boldsymbol{\rho}^{\textrm{group}}_{l,m}= \big[[\boldsymbol{\rho}^{\textrm{grid}}_l]_{m},[\boldsymbol{\rho}^{\textrm{grid}}_l]_{m+M},\cdots,[\boldsymbol{\rho}^{\textrm{grid}}_l]_{m+(I-1)M}\big]$.
The problem in \eqref{Local Grid Target Sensing} is a group sparse signal recovery problem that can be efficiently, albeit suboptimally, solved using existing sparse recovery algorithms. In this work, we adopt the OMP algorithm \cite{pati1993orthogonal}, following the approach in Sec. \ref{Sec:Grid-Based Sensing Algorithm}.

\textcolor{black}{Similarly, the non-cooperative grid-free sensing problem for gateway $l$ can be formulated as
\begin{subequations}
\begin{align}
\min_{\{\mathbf{q}_{k,l}\}_{k\in\mathcal{K}},\{\boldsymbol{\rho}^{\textrm{cont}}_{k,l}\}_{k\in\mathcal{K}}} ~~ &\big\|\mathbf{y}_l-\sum_{k\in\mathcal{K}}\mathbf{A}^{\textrm{cont}}_l(\mathbf{q}_{k,l}) \boldsymbol{\rho}^{\textrm{cont}}_{k,l}\big\|_2^2 \\
\textrm{s.t.} ~~~~~~~~~~\,\,&\mathbf{q}_{k,l} \in \mathbb{R}^3,\,\boldsymbol{\rho}^{\textrm{cont}}_{k,l}\in\mathbb{C}^{I},~~k\in\mathcal{K},
\end{align}
\end{subequations}
where $\boldsymbol{\rho}^{\textrm{cont}}_{k,l} = [\rho_{1,k,l},\cdots,\rho_{I,k,l}]^T$. This non-convex problem can be efficiently solved by applying the PSO algorithm, as in Sec. \ref{Sec:Grid-Free Sensing Algorithm}. For brevity, the detailed derivations of both non-cooperative sensing algorithms are omitted here. After solving either the grid-based or grid-free sensing problem, each gateway independently obtains $K$ local target position estimates $\hat{\mathcal{Q}}_l \triangleq  
\{\hat{\mathbf{q}}^{\textrm{tar}}_{k,l}\}_{k\in\mathcal{K}}$ from its own observation.}


\subsection{Data Association}\label{Sec:Data Association}
In the proposed distributed sensing framework, each gateway constructs a {\em candidate} set of target locations based on the non-cooperative sensing framework. After this, each gateway forwards the candidate set $\hat{\mathcal{Q}}_l$ to the CU for data fusion. The main challenge is how to match the estimates from different gateways to the right target. This is because each gateway provides $K$ potential target positions, but their association with a specific target is unknown. 

For accurate multi-target sensing, we devise a data association process to match the estimates (potentially) associated with the same target. For convenience, we define $\mathcal{C}_k$ as the set of location estimates for target $k$. The number of elements in $\mathcal{C}_k$ is given by $|\mathcal{C}_k|=L$, since each gateway contributes exactly one estimate. This also implies that if $\hat{\mathbf{q}}^{\textrm{tar}}_{k,l}\in \mathcal{C}_k$, then $\hat{\mathbf{q}}^{\textrm{tar}}_{k',l} \notin \mathcal{C}_k$ for any $k \neq k'$. Under these constraints, our goal is to find $\mathcal{C}_1,\cdots,\mathcal{C}_K$ that minimize the estimation error. The 2D structure of the UPA can provide precise DoA information, but not range information. Thus, the geometry of target $k$'s possible location as estimated by gateway $l$ is represented by a line passing through the gateway and its estimated position. Based on this geometry, we define the squared Euclidean distance between the line passing through two points $\mathbf{a},\mathbf{b}\in \mathbb{R}^3$ and a point $\mathbf{c}\in \mathbb{R}^3$ as
\begin{align}
f(\mathbf{a},\mathbf{b},\mathbf{c})=(\mathbf{c}\!-\!\mathbf{a})^T\left(\mathbf{I}_3-\frac{(\mathbf{b}-\mathbf{a})(\mathbf{b}-\mathbf{a})^T}{\|\mathbf{b}-\mathbf{a}\|^2}\right)(\mathbf{c}\!-\!\mathbf{a}).
\end{align}
Motivated by this, we formulate the data association problem to minimize the squared distance error, as follows:
\begin{subequations}
\label{Data association problem}
\begin{align}
&\min_{\mathcal{C}_1,\cdots,\mathcal{C}_K} ~~ \sum_{k\in\mathcal{K}}\sum_{l\in\mathcal{L}}\sum_{g\in\mathcal{L}\setminus\left\{l\right\}}f(\mathbf{q}^{\textrm{gat}}_{l},\mathbf{c}_{k,l},\mathbf{c}_{k,g})\\
&\,\,\,\,\,\,\,\,\textrm{s.t.}\,\,\,\,\,\,\,\,\,\, |\mathcal{C}_k|=L,~~k\in\mathcal{K}, \label{Data association problem constraint}\\
&\,\,\,\,\,\,\,\,\,\,\,\,\,\,\,\,\,\,\,\,\,\,\,\,\, |\mathcal{C}_k \cap \hat{\mathcal{Q}}_l |=1 ,~~k\in\mathcal{K},\,l\in\mathcal{L},\\
&\,\,\,\,\,\,\,\,\,\,\,\,\,\,\,\,\,\,\,\,\,\,\,\,\, \mathop{\scalebox{1.3}{$\cup$}}\limits_{k\in\mathcal{K}} \mathcal{C}_k = \mathop{\scalebox{1.3}{$\cup$}}\limits_{l\in\mathcal{L}} \hat{\mathcal{Q}}_l,
\end{align}
\end{subequations}
where $\mathbf{c}_{k,l}$ is the element of $\mathcal{C}_k$ corresponding to gateway $l$. This data association problem is a well-known $L$-assignment problem. For the case of $L=2$, it simplifies to a bipartite matching problem, which can be solved by the Hungarian algorithm \cite{kuhn1955hungarian}. The Hungarian algorithm finds an optimal one-to-one correspondence between two sets by iteratively updating a cost matrix. Due to its polynomial-time complexity, it efficiently provides an optimal solution for two-dimensional matching problems. However, when $L \geq 3$, the problem is known to be NP-hard \cite{frieze1983complexity}, since the bipartite structure no longer holds and the number of possible assignments increases exponentially with $L$. Practical multi-satellite networks generally consist of three or more gateways, thereby making the problem intractable. To address this issue, we alternatively solve problem \eqref{Data association problem} by sequentially applying the Hungarian algorithm. \textcolor{black}{This method does not guarantee a global optimal solution to problem \eqref{Data association problem}. However, it efficiently finds a suboptimal assignment by avoiding the prohibitive computational complexity of exhaustive search.} 

{\setlength{\textfloatsep}{1pt}
\begin{algorithm}[t] 
    \caption{Clustering of Non-Cooperative Estimates}\label{alg:data-association}
    {
    \small
    {\begin{algorithmic}[1]
        \REQUIRE $\{\hat{\mathcal{Q}}_l\}_{l\in\mathcal{L}}$, $\{\mathbf{q}^{\textrm{gat}}_{l}\}_{l\in\mathcal{L}}$ and $L$.
        \ENSURE $\{\mathcal{C}_k\}_{k\in \mathcal{K}}$.
        \STATE Initialize $\mathcal{C}_{1}^{(1)} =\{\hat{\mathbf{q}}^{\textrm{tar}}_{1,1}\}$, $\mathcal{C}_{2}^{(1)} =\{\hat{\mathbf{q}}^{\textrm{tar}}_{2,1}\}$, $\cdots$, $\mathcal{C}_{K}^{(1)} =\{\hat{\mathbf{q}}^{\textrm{tar}}_{K,1}\}$.
        \FOR{$g=2:L$}
        \STATE Formulate the cost matrix $\mathbf{C}^{(g)}$ as in \eqref{cost matrix}.
        \STATE Apply the Hungarian algorithm to the cost matrix.
        \STATE $\mathcal{C}_{k}^{(g)} \leftarrow \mathcal{C}_{k}^{(g-1)} \cup \{\mathbf{c}_{k,g}\}$.
        \ENDFOR
        \STATE \textbf{Return} $\mathcal{C}_1 = \mathcal{C}_{1}^{(L)},\mathcal{C}_2 = \mathcal{C}_{2}^{(L)},\cdots,\mathcal{C}_K = \mathcal{C}_{K}^{(L)}$.
    \end{algorithmic}}
    }
\end{algorithm}
}

For solving the data association problem, we first initialize the mapping solution $\mathcal{C}_1,\cdots,\mathcal{C}_K$ by assigning one element from the first gateway’s estimate $\hat{\mathcal{Q}}_1$ to each $\mathcal{C}_k$, i.e., $\mathcal{C}_{1}^{(1)} =\{\hat{\mathbf{q}}^{\textrm{tar}}_{1,1}\}$, $\mathcal{C}_{2}^{(1)} =\{\hat{\mathbf{q}}^{\textrm{tar}}_{2,1}\}$, $\cdots$, $\mathcal{C}_{K}^{(1)} =\{\hat{\mathbf{q}}^{\textrm{tar}}_{K,1}\}$. At the $g$-th iteration, our objective is to assign $\hat{\mathbf{q}}^{\textrm{tar}}_{1,g},\hat{\mathbf{q}}^{\textrm{tar}}_{2,g},\cdots,\hat{\mathbf{q}}^{\textrm{tar}}_{K,g}$ to $\mathcal{C}^{(g-1)}_1,\mathcal{C}^{(g-1)}_2,\cdots,\mathcal{C}^{(g-1)}_K$, thereby determining $\mathbf{c}_{1,g},\mathbf{c}_{2,g},\cdots,\mathbf{c}_{K,g}$ so as to minimize the geometric inconsistency with the mapping obtained in the previous iterations. To this end, we introduce a cost matrix $\mathbf{C}^{(g)}\in\mathbb{R}^{K\times K}$, whose $(k,k')$-th element is defined as
\begin{align}
\Big[\mathbf{C}^{(g)}\Big]_{k,k'}=\sum_{l=1}^{g-1}f(\mathbf{q}^{\textrm{gat}}_{g},\hat{\mathbf{q}}^{\textrm{tar}}_{k,g},\mathbf{c}_{k',l}).
\label{cost matrix}
\end{align}
We then apply the Hungarian algorithm to $\mathbf{C}^{(g)}$, from which the assignments $\mathbf{c}_{1,g},\cdots, \mathbf{c}_{K,g}$ are obtained.
Consequently, the original problem can be solved by sequentially applying the Hungarian algorithm until the constraint \eqref{Data association problem constraint} is satisfied. The detailed procedure of the proposed data association is presented in Algorithm \ref{alg:data-association}. \textcolor{black}{Since the proposed algorithm depends on the ordering of the gateways, assigning higher priority to more reliable gateways can further improve the data association performance. However, in this work, the reliability of each gateway is assumed to be unknown. Therefore, a fixed gateway ordering is adopted.} 

{\setlength{\textfloatsep}{1pt}
\begin{table*}[t] 
{\small 
\centering
\caption{\textcolor{black}{Computational Complexity Comparison of Sensing Frameworks}}
\color{black}
\renewcommand{\arraystretch}{1.2}
\begin{tabular}{
    | >{\centering\arraybackslash}m{3.5cm} |
      >{\centering\arraybackslash}m{10.2cm} |
      >{\centering\arraybackslash}m{3.1cm} |
}
\hline
\textbf{Method} & \textbf{Total computational complexity} & \textbf{Dominant scaling} \\
\hline
\noalign{\vskip 1mm}
\hline
Centralized grid-based & $\mathcal{O}(KMTIL^2 + K^3TI^2L^3 + K^4I^3L^3)$ & $\mathcal{O}(KMTIL^2)$ \\\hline
Centralized grid-free & $\mathcal{O}(KPN(TI^2L^3+I^3L^3) + K^3TI^2L^3 + K^4I^3L^3)$ & $\mathcal{O}(KPNTI^2L^3)$ \\\hline
Distributed grid-based & $\mathcal{O}(L(KMTI+K^3I^2T+K^4I^3)+K^2L^2+K^3L+KL)$ &  $\mathcal{O}(KMTIL)$\\\hline
Distributed grid-free & $\mathcal{O}(L(KPN(TI^2+I^3)+K^3TI^2+K^4I^3)+K^2L^2+K^3L+KL)$ & $\mathcal{O}(KPNTI^2L)$\\
\hline
\end{tabular}
\label{Table: Computational Complexity}
\vspace{-3mm}
}
\end{table*}  
}

\subsection{Data Fusion}\label{Sec:Data Fusion}
After the data association described above, the CU determines the final estimate for the position of target $k$ from the obtained solution $\mathcal{C}_k$. To this end, we formulate the distance error minimization problem:  
\begin{align}
\hat{\mathbf{q}}^{\textrm{tar}}_{k} = \underset{\mathbf{q}\in \mathbb{R}^3}{\arg\!\min}~\sum_{l\in\mathcal{L}}f(\mathbf{q}^{\textrm{gat}}_{l},\mathbf{c}_{k,l},\mathbf{q}).
\label{Data fusion problem}
\end{align}
To solve this problem, we first rewrite the squared distance as
\begin{align}
f(\mathbf{q}^{\textrm{gat}}_{l},\mathbf{c}_{k,l},\mathbf{q})\! = (\mathbf{q} - \mathbf{q}^{\textrm{gat}}_{l})^T\left(\mathbf{I}_3-\mathbf{d}_{k,l}\mathbf{d}_{k,l}^T\right)(\mathbf{q} - \mathbf{q}^{\textrm{gat}}_{l}),
\end{align}
where $\mathbf{d}_{k,l} = (\mathbf{c}_{k,l}-\mathbf{q}^{\textrm{gat}}_{l})/ \|\mathbf{c}_{k,l}-\mathbf{q}^{\textrm{gat}}_{l}\|$. For any vector $\mathbf{q}\in\mathbb{R}^3$, it follows that $\mathbf{q}^T(\mathbf{I}_3-\mathbf{d}_{k,l}\mathbf{d}_{k,l}^T)\mathbf{q}=\|\mathbf{q}\|^2\sin^2\alpha$,
where $\alpha \in [0, \pi]$ denotes the angle between $\mathbf{q}$ and $\mathbf{d}_{k,l}$. Since $\sin^2\alpha \geq 0$ for all $\alpha$, we have $\mathbf{q}^T(\mathbf{I}_3-\mathbf{d}_{k,l}\mathbf{d}_{k,l}^T)\mathbf{q}\geq 0,$
which implies that $\mathbf{I}_3-\mathbf{d}_{k,l}\mathbf{d}_{k,l}^T$ is a positive semidefinite matrix.
Thus, the problem \eqref{Data fusion problem} is an unconstrained quadratic program. According to the first-order optimality condition, the optimal solution can be obtained by setting the derivative to zero:
\begin{align}
\frac{\partial }{\partial \mathbf{q}}	\Bigg[\sum_{l\in\mathcal{L}}f(\mathbf{q}^{\textrm{gat}}_{l},\mathbf{c}_{k,l},\mathbf{q})\Bigg]\Bigg|_{\mathbf{q}=\hat{\mathbf{q}}^{\textrm{tar}}_{k}} &= 2\sum_{l\in\mathcal{L}}\mathbf{D}_{k,l}(\hat{\mathbf{q}}^{\textrm{tar}}_{k} - \mathbf{q}^{\textrm{gat}}_{l})\nonumber\\
&=\mathbf{0}_3,
\end{align}
where $\mathbf{D}_{k,l}=\mathbf{I}_3-\mathbf{d}_{k,l}\mathbf{d}_{k,l}^T$. As a result, the estimated position of target $k$ under the distributed framework is obtained as
\begin{align}
\hat{\mathbf{q}}^{\textrm{tar}}_{k} = \left(\sum_{l\in\mathcal{L}}\mathbf{D}_{k,l}\right)^{-1}\left(\sum_{l\in\mathcal{L}}\mathbf{D}_{k,l}\mathbf{q}^{\textrm{gat}}_{l}\right). \label{Distributed sensing result}
\end{align}

{\color{black}
\subsection{Scalability and CRLB Analysis}
}
\textcolor{black}{
For a wide sensing region requiring a large search space, the distributed framework can be extended by partitioning the sensing region into multiple subregions. Specifically, each gateway applies the sensing algorithm only to the corresponding subregion in each sensing interval. This reduces the search complexity compared with processing the entire sensing region at once. The target estimates obtained from different gateways and sensing intervals can then be associated using the Hungarian algorithm before fusion. This extension may reduce sensing accuracy due to the loss of joint processing over the entire sensing region. Nevertheless, it improves the scalability of the proposed distributed framework, facilitating flexible expansion of the sensing region.}

\textcolor{black}{
We also discuss the CRLB analysis for the proposed distributed framework. The CRLB of the non-cooperative framework can be derived by following the same logic as in Sec. \ref{Sec:Theoretical Analysis}. However, such a derivation is not straightforward for the distributed framework because its final estimates are obtained from local sensing results, making it difficult to formulate a unified likelihood function. To address this issue, we focus on the fact that information loss is inevitable during local processing. This suggests that the centralized CRLB can serve as a theoretical limit for the proposed distributed framework.}

\subsection{Analysis for Complexity and Signaling Overhead}\label{Complexity analysis:distributed-sensing}
\textcolor{black}{The computational complexities of the proposed sensing frameworks are summarized in Table \ref{Table: Computational Complexity}, where the dominant scaling is obtained under the practical sensing assumptions that $K \ll M$, $K \ll PN$, and $I \ll T$. As shown in Table \ref{Table: Computational Complexity}, the grid-free methods have higher computational complexity than the grid-based ones due to the iterative nature of the PSO algorithm. This makes the grid-based methods advantageous for real-time applications. Meanwhile, the grid-free methods support high-resolution sensing by directly searching the continuous space at the cost of increased computational complexity. Joint processing entails higher complexity for the centralized frameworks than for the distributed frameworks.} 

The distributed framework requires much lower fronthaul overhead than the centralized framework, since only local sensing results are forwarded to the CU instead of raw observations. Specifically, in the centralized sensing framework, each gateway forwards its raw observation vector $\mathbf{y}_l$ to the CU, which is equivalent to transmitting $T$ complex-valued observations per gateway in each sensing period. In contrast, in the distributed framework, each gateway only transmits the estimated candidate set $\hat{\mathcal{Q}}_l$, which consists of only $K$ three-dimensional target position estimates. This leads to a considerable reduction in fronthaul overhead, which makes the distributed framework more suitable for practical multi-satellite ISAC systems with limited fronthaul. 

\section{Numerical Results}\label{Sec:Numerical Results}
In this section, we evaluate the superiority of the proposed frameworks via simulations. \textcolor{black}{For the simulations, we consider a wide-area sensing task for the detection of high-altitude platforms.} The satellites are located at $\mathbf{q}^{\textrm{sat}}_{1} = [50,  50, 600]^T \textrm{ km}$ and $\mathbf{q}^{\textrm{sat}}_{2} = [-50,  -50, 600]^T \textrm{ km}$. \textcolor{black}{The gateways are deployed\footnote{\textcolor{black}{The gateways are deployed on the same plane, which makes the fusion less well-conditioned in the vertical dimension. However, non-coplanar deployment may be infeasible in some practical satellite networks due to geographical and infrastructural constraints. For this reason, in this work, we evaluate the sensing frameworks without geometric gains from different gateway altitudes.}} at $\mathbf{q}^{\textrm{gat}}_{1} = [10,  10, 0]^T \textrm{ km}$, $\mathbf{q}^{\textrm{gat}}_{2} = [10,  -10, 0]^T \textrm{ km}$, $\mathbf{q}^{\textrm{gat}}_{3} = [-10,  10, 0]^T \textrm{ km}$, and $\mathbf{q}^{\textrm{gat}}_{4} = [-10,  -10, 0]^T \textrm{ km}$.} The ground UEs are randomly placed within a circular area of $50 \textrm{ km}$ diameter centered at each satellite, with a minimum inter-UE distance of $10 \textrm{ km}$. We also assume that the airborne targets are uniformly distributed within a circular region of $10 \textrm{ km}$ diameter centered at the origin, and their altitudes are between $17 \textrm{ km}$ and $20 \textrm{ km}$ \cite{karapantazis2005broadband}. We do not consider targets with reflection coefficient magnitudes below 3 because their echo power is too weak for reliable sensing at the assumed SNR level. This sensing region is discretized into $M=324$ points with a spacing of $1 \textrm{ km}$. For target sensing, the communication and sensing symbols are independently generated with unit norm and random phases uniformly distributed over $[0,2\pi)$, and the mapping function is defined as $\psi(t)=((t-1) \textrm{ mod } M)+1$. The noise power is set to $\sigma_{i,u}^2=\zeta_l^2=N_{0}B$ with the noise power spectral density $N_{0} = -174 \textrm{ dBm/Hz}$ and downlink channel bandwidth $B = 5 \textrm{ MHz}$ \cite{3gppR1_2401845}. The remaining parameters follow 3GPP R1-2401845 \cite{3gppR1_2401845} and existing studies \cite{park2025bistatic},\cite{yoo2024cache},\textcolor{black}{\cite{raju2011rcs}}, as summarized in Table \ref{Simulation Parameters}. \textcolor{black}{Under this setup, the average received sensing SNR of each satellite-gateway bistatic path is approximately $13.26 \textrm{ dB}$ for a target at the center of the sensing region with $P_{i}^{\textrm{r}}=1 \textrm{ W}$ and target-aligned beamforming.}

\textcolor{black}{We consider the RMSE as the sensing performance metric, which is defined as $\sqrt{\frac{1}{K}\sum_{k \in \mathcal{K}}\|\mathbf{q}^{\textrm{tar}}_{k} - \hat{\mathbf{q}}^{\textrm{tar}}_{k}\|^2_2}$. The communication performance is also evaluated via the average SINR and average transmit power, which are calculated as $\frac{1}{TU}\sum_{t\in\mathcal{T}}\sum_{i\in\mathcal{I}}\sum_{u\in\mathcal{U}_i} \textrm{SINR}^{\textrm{ue}}_{i,u}(t)$ and $\frac{1}{TU}\sum_{t\in\mathcal{T}}\sum_{i\in\mathcal{I}}\sum_{u\in\mathcal{U}_i} p_{i,u}^{\textrm{c}}(t)$, respectively.} All simulation results are averaged over 1000 random samples.

{\setlength{\textfloatsep}{1pt}
\begin{table}[t]
{\small
\centering
    \vspace{-10pt}
    \caption{Simulation Parameters}
    \label{Simulation Parameters}
    \centering
    \vspace{-4pt}
    \begin{tabular}{cccc} 
        \\[-1.8ex]\hline 
        \hline \\[-1.8ex] 
        \multicolumn{1}{c}{\textbf{Parameter}} & \multicolumn{1}{c}{\textbf{Value}} & \multicolumn{1}{c}{\textbf{Parameter}} & \multicolumn{1}{c}{\textbf{Value}}\\
        \hline \\[-1.8ex] 
        $\lambda$ & 0.15 m & $N_{x}^{\textrm{sat}} \times N_{y}^{\textrm{sat}}$ & $26\times26$\\
        $A^{\textrm{sat}}_{\textrm{Tx}}$ & 30 dBi & $N_{x}^{\textrm{gat}} \times N_{y}^{\textrm{gat}}$ & $32\times32$ \cite{park2025bistatic}\\
        $A^{\textrm{gat}}_{\textrm{Rx}}$ & 30 dBi & $\kappa$ & 30 dB \cite{yoo2024cache}\\
        $A^{\textrm{ue}}_{\textrm{Rx}}$ & -5.5 dBi & $U_i$ & 5\\
        $\gamma_{i,k,l}$ & 10 dBsm\textcolor{black}{\cite{raju2011rcs}}&  $\tau_{\textrm{c}}$ & -10 dB \\
        \\[-1.8ex]\hline  
        \hline \\[-1.8ex] 
    \end{tabular}
    \vspace{-4pt}
    }
\end{table}
}

For performance comparisons, the following benchmark schemes are considered.
\begin{itemize}
\item \textbf{OMP-CEN} and \textbf{OMP-DIS}:
The grid-based centralized and distributed sensing frameworks described in
Secs. \ref{Sec:Grid-Based Sensing Algorithm} and
\ref{Sec:Distributed Sensing Framework}, respectively. OMP-DIS solves the non-cooperative sensing problem in \eqref{Local Grid Target Sensing} at each gateway.
\item \textbf{OMP-NC}:
The non-cooperative sensing framework that only uses the observation from the first gateway and then applies the OMP method.
\item \textcolor{black}{\textbf{PSO-CEN}, \textbf{PSO-DIS}, and \textbf{PSO-NC}: The grid-free centralized, distributed, and non-cooperative sensing frameworks, respectively. These frameworks are counterparts of OMP-CEN, OMP-DIS, and OMP-NC, respectively, with the PSO algorithm adopted instead of OMP. For all PSO-based frameworks, we set $P=50$, $N=80$, $w_{\max}=0.8$, $w_{\min}=0.4$, and $c_1=c_2=1.5$.}
\item \textbf{MUSIC-CEN} and \textbf{MUSIC-NC}:     
The classical MUSIC algorithm \cite{schmidt1986multiple} implemented with one satellite and either four
gateways (centralized) or a single gateway (non-cooperative), where the sensing power is
set to $P_i^{\textrm{r}} I$ and the number of time slots is $10M$. In the centralized case, the MUSIC pseudo-spectrum is defined as
$1/\sum_{l\in \mathcal{L}}
(\mathbf{v}^{\textrm{gat-grid}}_{l,m})^{H}
\mathbf{U}_{l}\mathbf{U}_{l}^{H}
\mathbf{v}^{\textrm{gat-grid}}_{l,m},
$ where $\mathbf{U}_{l}$ denotes the noise subspace obtained from the sample
covariance matrix of the received signals at gateway $l$.
\item \textbf{CoSaMP-CEN} and \textbf{CoSaMP-DIS}:    
Variants of the grid-based centralized and distributed sensing frameworks, where compressive sampling matching pursuit (CoSaMP) \cite{needell2009cosamp} is applied for sparse signal recovery, instead of OMP. 
\item \textbf{OMP-NC + K-means}:
A distributed sensing framework in which $K$-means clustering \cite{macqueen1967multivariate} is applied for data association instead of the Hungarian algorithm.
At each iteration, $K$-means clustering is applied to the local sensing results to extract one target location.
\item \textbf{Grid bound}:
The RMSE between each true target location and its nearest grid point,
which is a performance bound for grid-based search methods.
\item \textcolor{black}{\textbf{RCRLB}: 
The root CRLB (RCRLB) for the target positions, expressed as $\sqrt{\textrm{tr}\left(\textrm{CRLB}(\boldsymbol{\vartheta})\right)/K}$,
where $\textrm{CRLB}(\boldsymbol{\vartheta})$ is defined in \eqref{eq:CRLB}, and $\textrm{tr}(\textrm{CRLB}(\boldsymbol{\vartheta}))$ is the sum of the diagonal entries of $\textrm{CRLB}(\boldsymbol{\vartheta})$. The RCRLB provides the minimum achievable root-mean-square error (RMSE), which is a theoretical benchmark for target estimation.}
\end{itemize}

\begin{figure}[t]
\captionsetup{labelfont={color=blue},font={color=blue}}
  \centering
  \includegraphics[width=5.95cm]{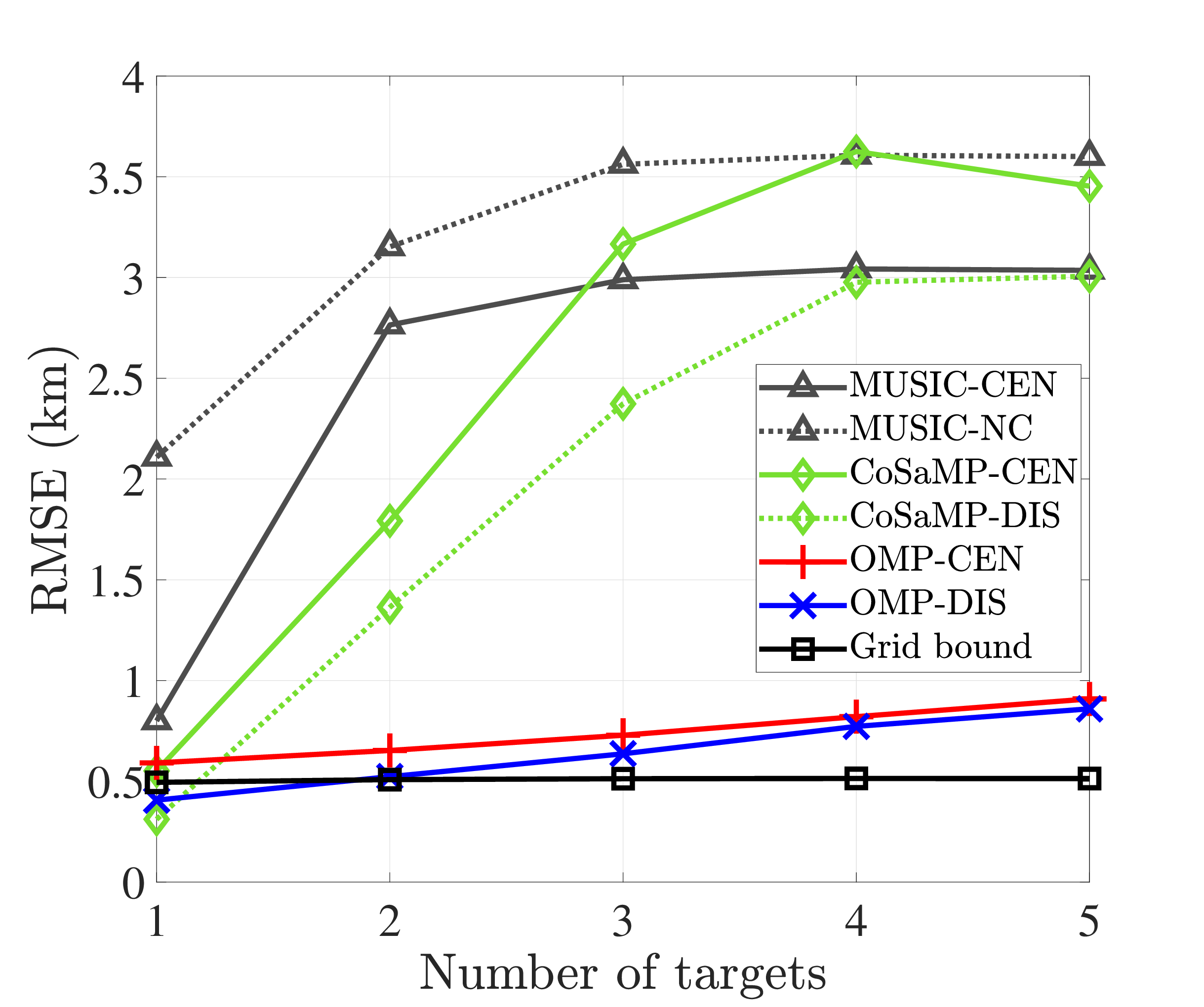}
    \caption{RMSE versus the number of targets ($T=M$ and $P_{i}^{\textrm{r}}=5 \textrm{ W}$).}
    \label{Fig:Target_Simulation}
    \vspace{-4mm}
\end{figure}

\begin{figure}[t]
\captionsetup{labelfont={color=blue},font={color=blue}}
  \centering
  \includegraphics[width=5.95cm]{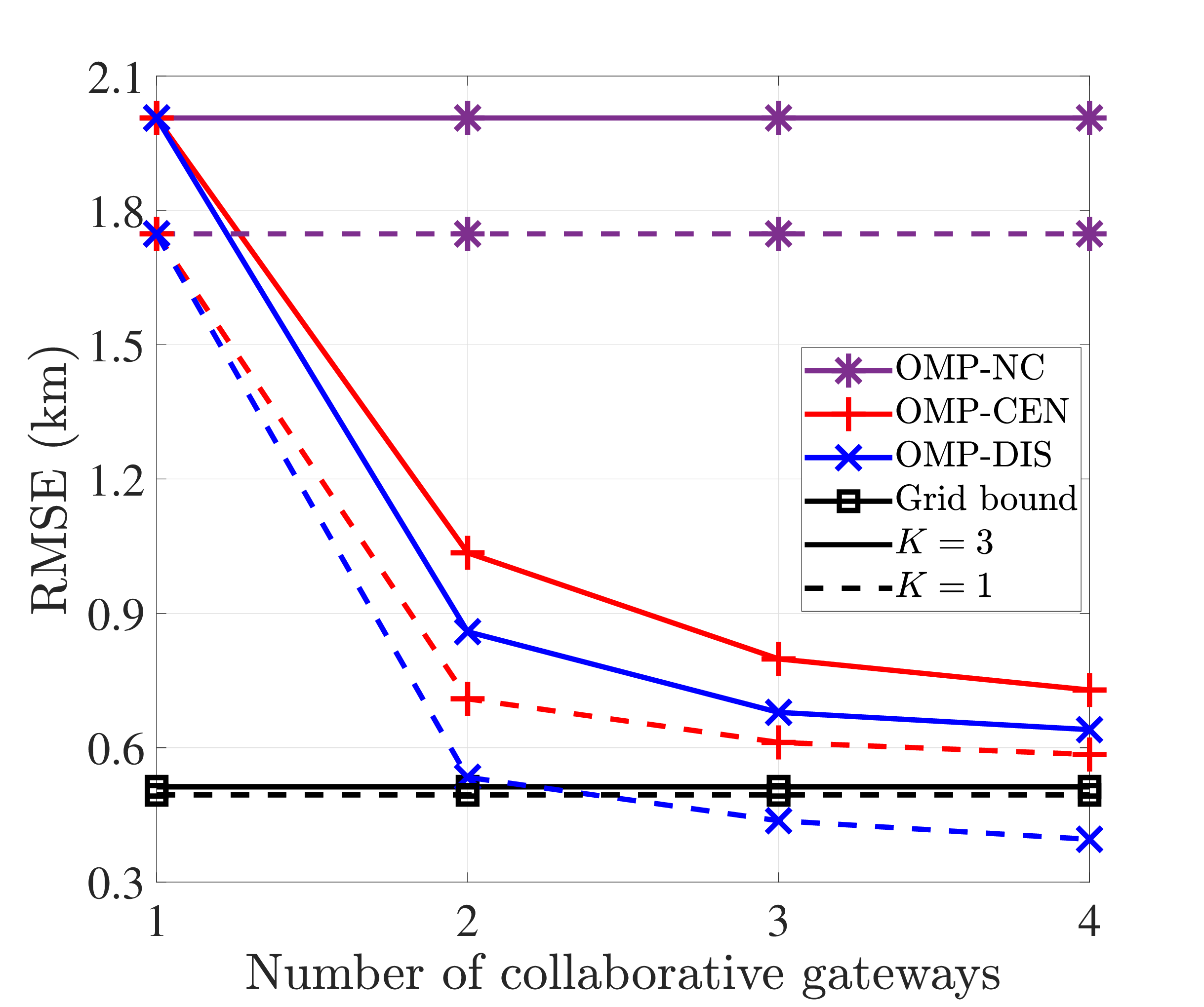}
    \caption{RMSE versus the number of collaborative gateways ($T=M$ and $P_{i}^{\textrm{r}}=1 \textrm{ W}$).}
    \label{Fig:Gateway_Simulation}
    \vspace{-4mm}
\end{figure}

Fig. \ref{Fig:Target_Simulation} plots the RMSE versus the number of targets with the grid-based sensing algorithms. The figure shows that both the proposed centralized and distributed frameworks outperform other baselines.
\textcolor{black}{The figure also shows that the distributed framework outperforms the centralized one under sufficient resources, despite requiring substantially lower fronthaul overhead and computational complexity. Specifically, in the single-target case, the distributed framework surpasses the grid bound, owing to its ability to resolve off-grid targets through data fusion.} 
On the contrary, the centralized framework is restricted to on-grid estimation and is therefore bounded by the grid error.

Compared to the compressed sensing benchmark, the OMP-based algorithm achieves more accurate sensing performance in multi-target scenarios. For CoSaMP, off-grid basis mismatch spreads target energy over coherent atoms, resulting in incorrect pruning and performance loss. The MUSIC algorithm can provide precise AoD information, but not range information. For this reason, centralized MUSIC achieves better performance than the non-cooperative one; however, it still shows a larger error compared to the proposed frameworks.


Fig. \ref{Fig:Gateway_Simulation} plots the RMSE versus the number of collaborative gateways. As shown in Fig. \ref{Fig:Gateway_Simulation}, the performance of the proposed grid-based frameworks improves significantly as more gateways cooperate for sensing. In contrast, the non-cooperative framework does not exhibit such performance gains, as it relies solely on individual sensing without cooperation. 

\begin{figure}[t]
\captionsetup{labelfont={color=blue},font={color=blue}}
    \centering 
    \subfigure[Grid-based frameworks.]
        {\includegraphics[width=5.95cm]{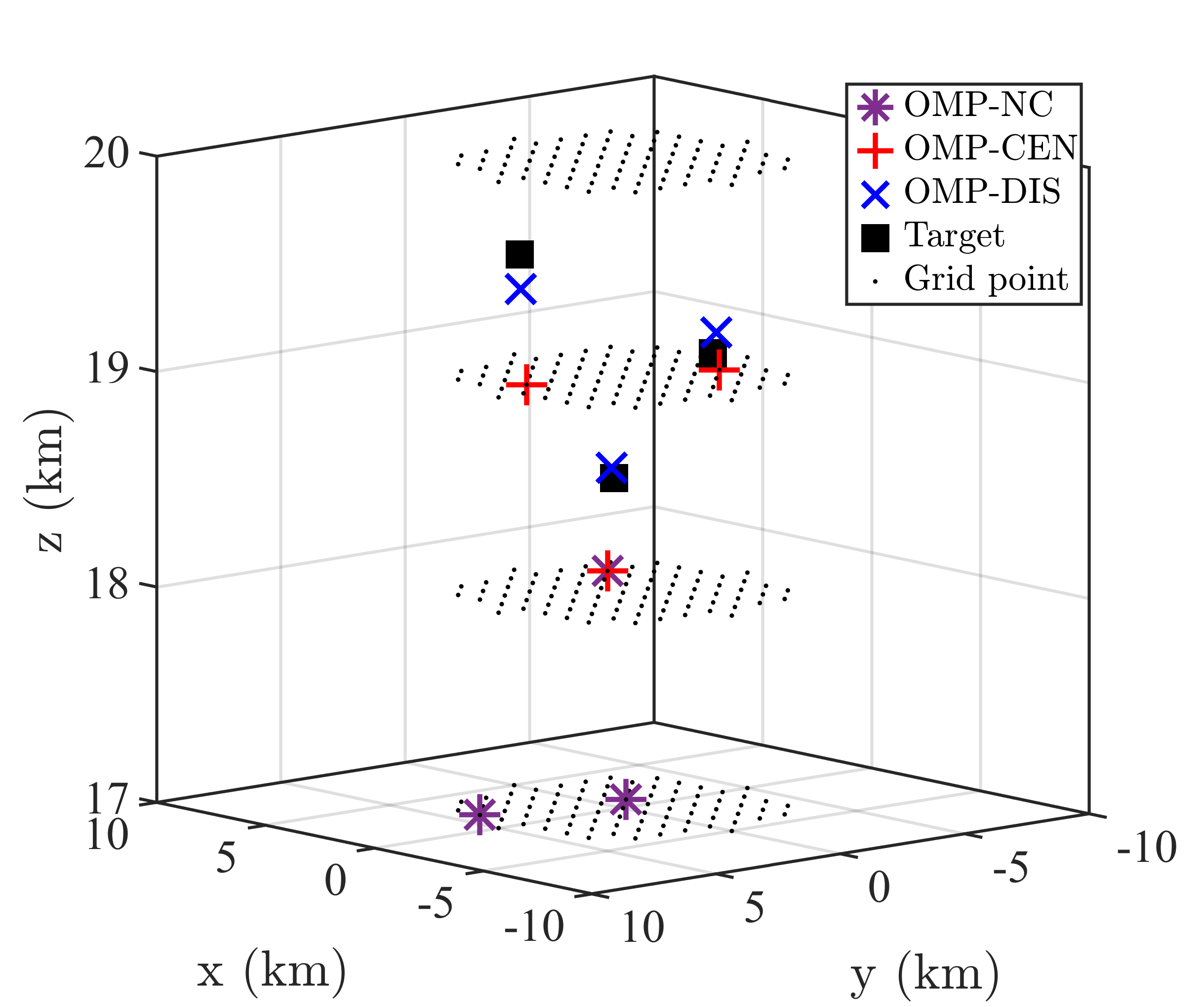}}
    \subfigure[Grid-free frameworks.]
    {\includegraphics[width=5.95cm]{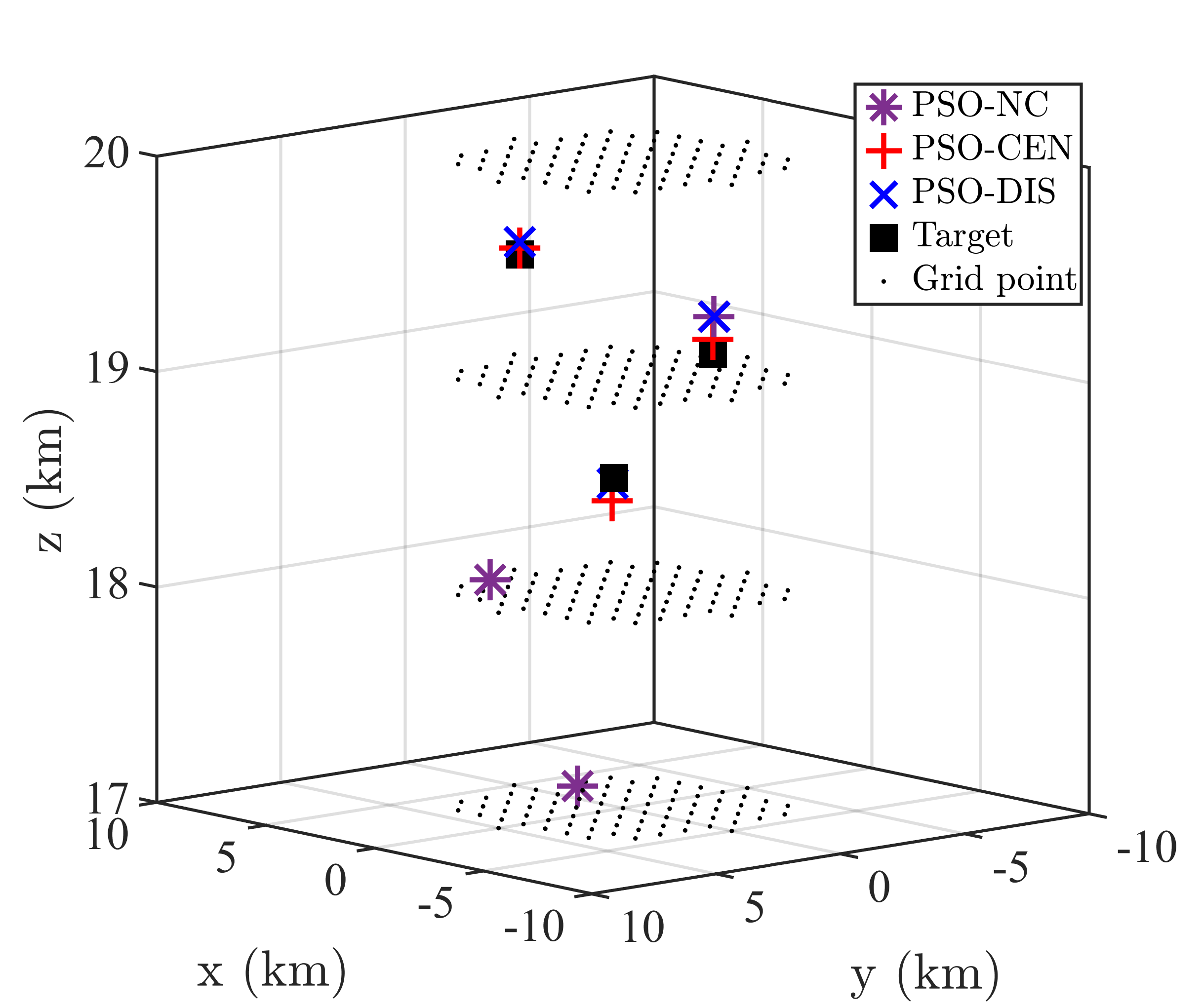}}
    \caption{Snapshots of the sensing results obtained from the proposed frameworks ($K=3$, $T=M$, and $P_{i}^{\textrm{r}}=1 \textrm{ W}$).} 
    \label{Fig: snapshot}
    \vspace{-4mm}
\end{figure}

Fig. \ref{Fig: snapshot} shows the snapshots of the sensing results of the proposed sensing frameworks. 
As shown in Fig. \ref{Fig: snapshot}, the non-cooperative sensing frameworks can provide accurate angular information; however, they exhibit poor performance in estimating the target range. This is because the two-dimensional structure of the UPA primarily captures angular information. For this reason, post-processing is required to compensate for the inaccurate range information, as done in the proposed distributed frameworks. Compared with non-cooperative sensing, both the centralized and distributed frameworks achieve improved estimation accuracy by combining raw observations or local sensing results from multiple gateways. Notably, even if the distributed frameworks fuse inaccurate estimates, they still attain a clear performance improvement over non-cooperative frameworks. This result verifies the effectiveness of the proposed data association and fusion mechanisms. \textcolor{black}{It can also be observed that OMP-CEN selects the nearest grid points to the target locations. By contrast, PSO-CEN provides more accurate sensing performance by directly searching the continuous space at the cost of increased computational complexity.}

\begin{figure}[t]
\captionsetup{labelfont={color=blue},font={color=blue}}
  \centering
  \includegraphics[width=5.95cm]{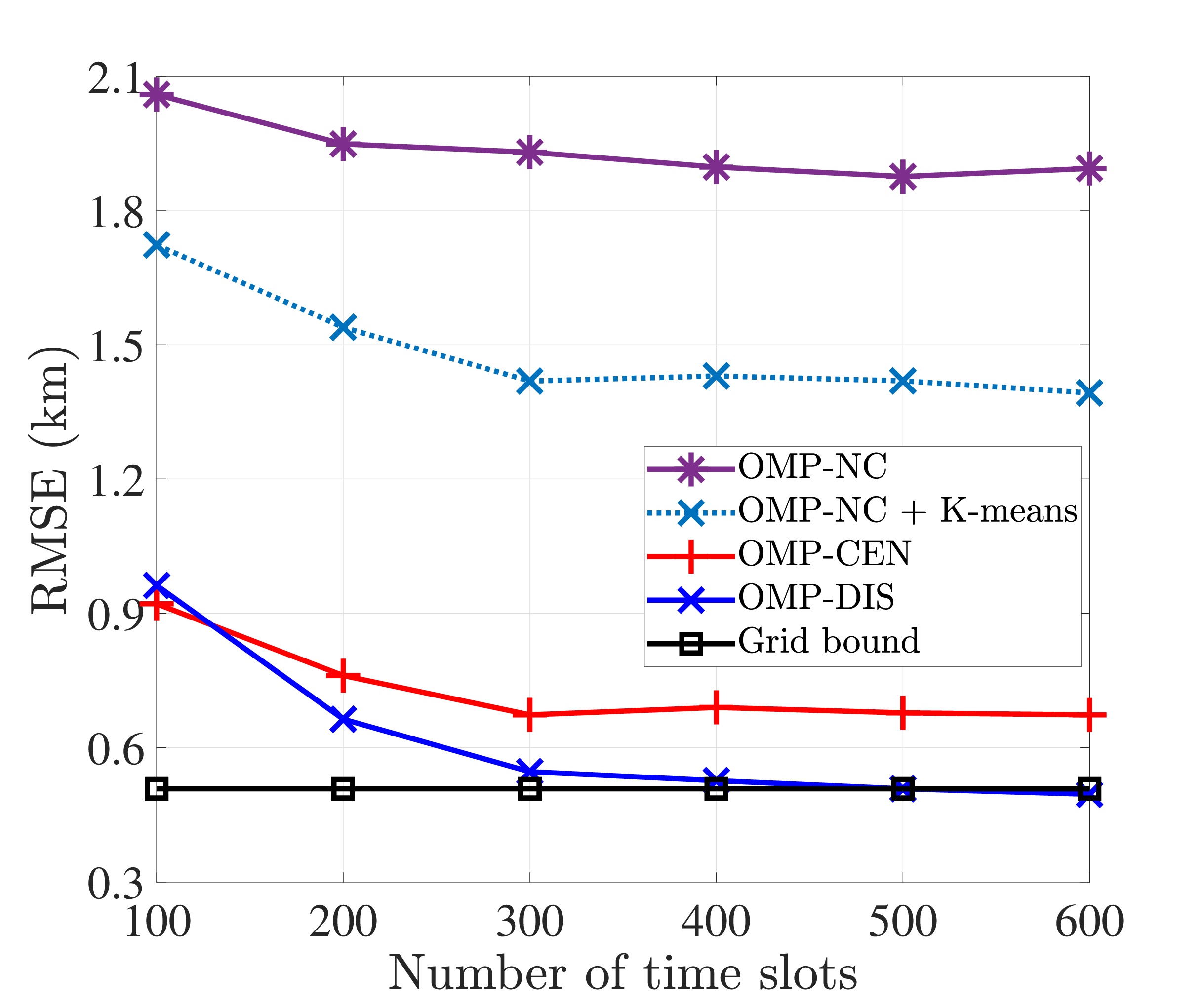}
    \caption{RMSE versus the number of time slots ($K=2$ and $P_{i}^{\textrm{r}}=1 \textrm{ W}$).}
    \label{Fig:Timeslot_Simulation}
    \vspace{-4mm}
\end{figure}

Fig. \ref{Fig:Timeslot_Simulation} depicts the RMSE versus the number of time slots for the grid-based sensing frameworks. 
We first observe that an increased number of time slots is beneficial for all schemes.
This is because more observations enhance measurement diversity, thereby improving
the accuracy of target estimation. Moreover, $K$-means clustering provides a noticeable improvement over the non-cooperative sensing scheme by grouping locally estimated target locations. However, this improvement is smaller than that from the proposed distributed framework. This is because $K$-means clustering is based on Euclidean distance, which is not suitable for the 2D UPA structure that mainly provides AoD information. In contrast, Algorithm \ref{alg:data-association} utilizes angular information via the Hungarian algorithm, which provides more effective clustering of local estimates. 
\textcolor{black}{We also observe that, when the number of time slots is insufficient, the centralized framework achieves a lower RMSE than the distributed framework due to its joint processing capability.}

\begin{figure}[t]
\captionsetup{labelfont={color=blue},font={color=blue}}
    \centering 
    \subfigure[$K=1$.]
        {\includegraphics[width=5.95cm]{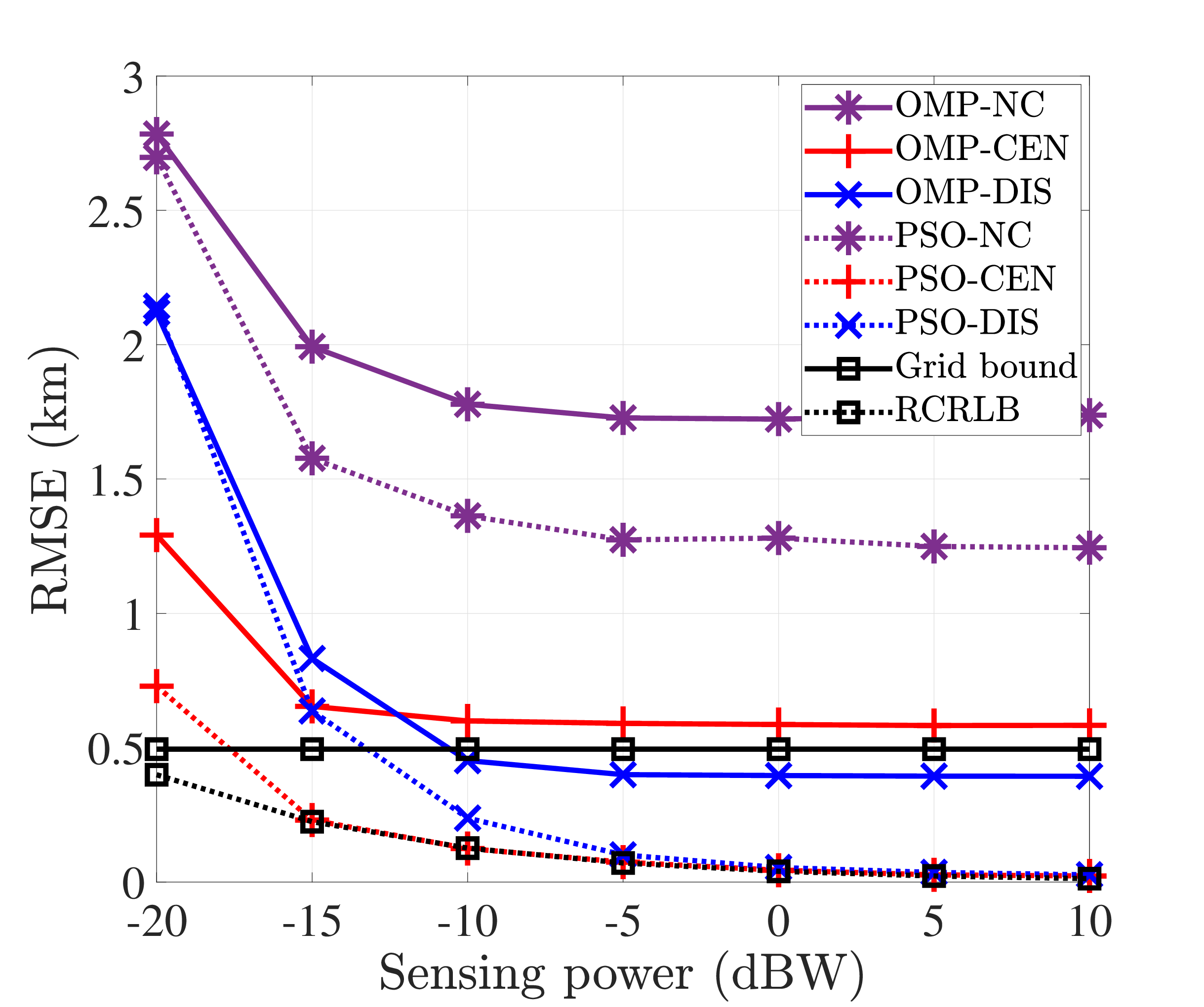}}
    \subfigure[$K=3$.]
    {\includegraphics[width=5.95cm]{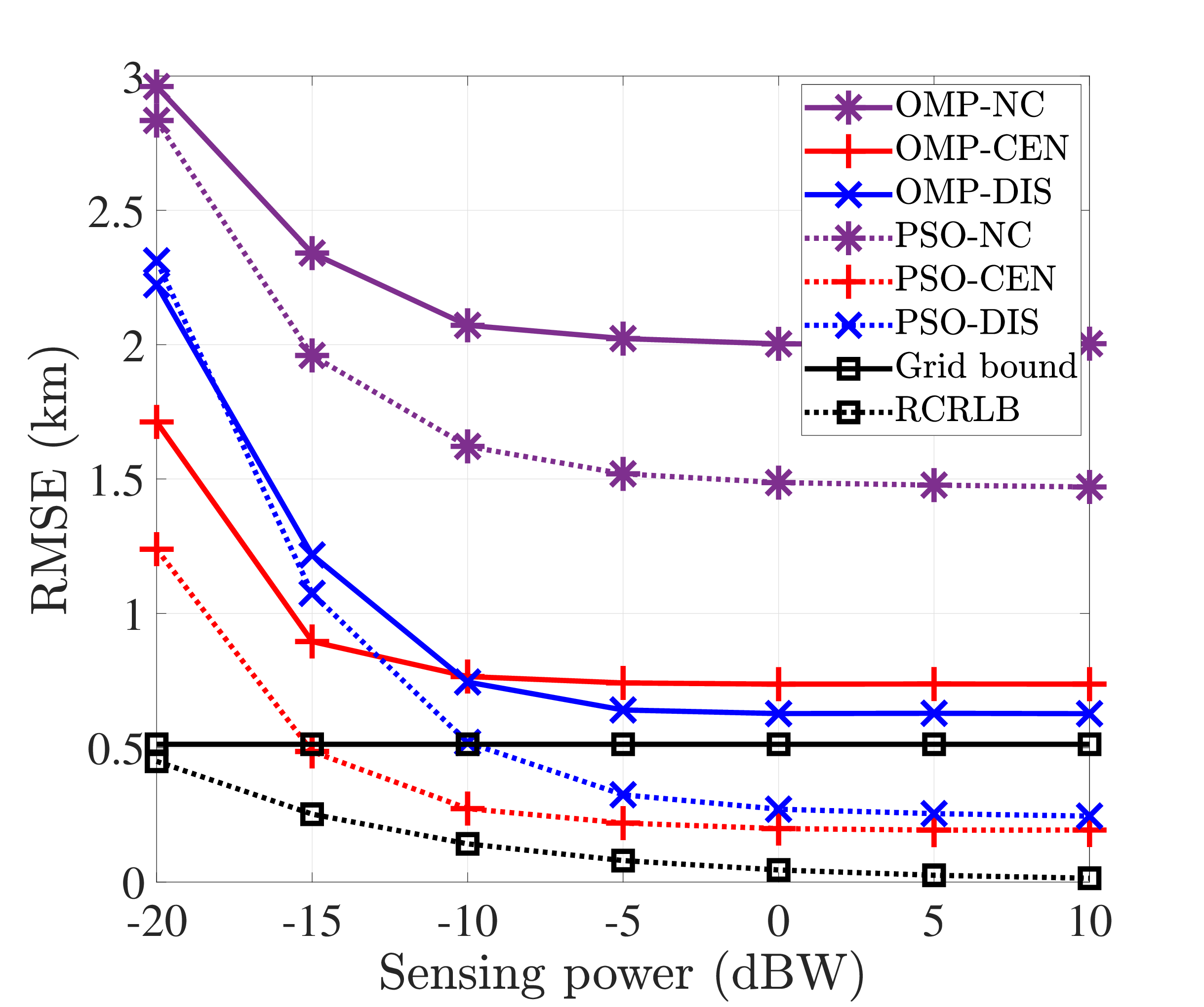}}
    \caption{RMSE versus the sensing power ($T=M$). } 
    \label{Fig:Sensing_Power_Simulation}
    \vspace{-4mm}
\end{figure}

Fig. \ref{Fig:Sensing_Power_Simulation} plots the RMSE versus the sensing power. \textcolor{black}{The grid-free frameworks achieve superior performance compared with the grid-based frameworks, as their search space is not confined to grid points. For the single-target case, the performance gap between the proposed cooperative grid-free frameworks and the RCRLB vanishes as the sensing power increases. This implies that they can achieve near-bound accuracy in the high sensing power regime. However, this gap is not negligible in the multi-target scenario because the superposition of multiple target echoes makes it more difficult to separate individual targets. The figure also shows that PSO-CEN consistently outperforms PSO-DIS across all regions. Nevertheless, the performance of PSO-DIS is comparable to that of PSO-CEN in the high sensing power regime. This observation validates the effectiveness of the proposed data association and fusion method.} 

\textcolor{black}{We next analyze the performance of the grid-based frameworks. It is observed that the centralized framework achieves the best performance in the low sensing power regime, as jointly processing raw observations is more effective under low SNR conditions. As the sensing power increases, the distributed framework improves significantly and surpasses the centralized one due to more reliable local estimation and effective data fusion.} For all schemes, the RMSE decreases with sensing power but saturates beyond a certain level, since structural factors such as grid resolution and basis mismatch constitute the dominant performance bottleneck rather than noise.

\begin{figure}[t]
\captionsetup{labelfont={color=blue},font={color=blue}}
  \centering
  \includegraphics[width=5.95cm]{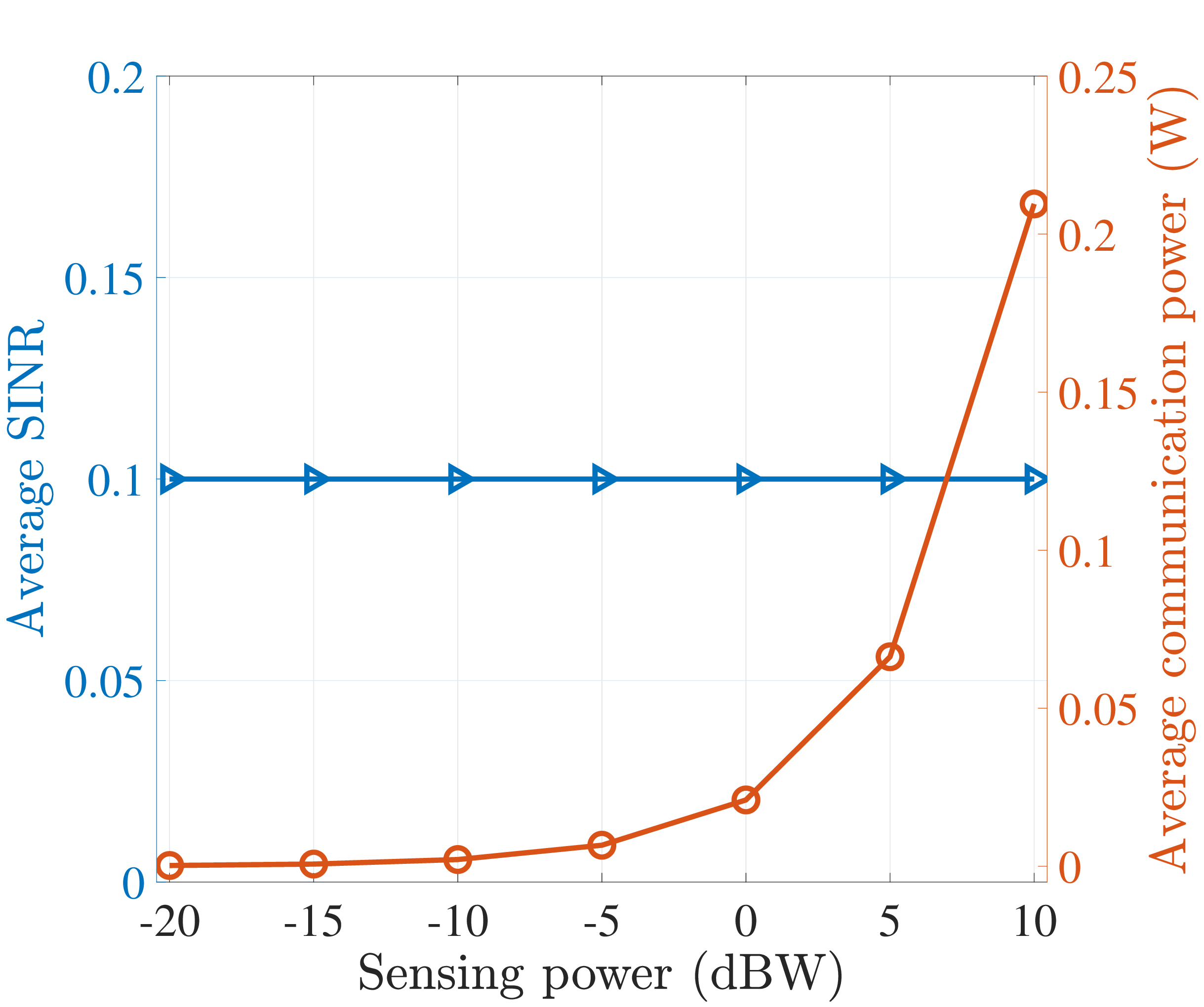}
    \caption{Average SINR and average communication power versus the sensing power ($T=M$).}
    \label{Fig:Sensing_Power_SINR_Simulation}
    \vspace{-4mm}
\end{figure}

\textcolor{black}{Fig. \ref{Fig:Sensing_Power_SINR_Simulation} plots the average SINR and average communication power versus the sensing power. The proposed power allocation method satisfies the SINR constraints regardless of the sensing power. However, the communication power consumption increases with the sensing power to compensate for the increased sensing interference. The results in Figs. \ref{Fig:Sensing_Power_Simulation} and \ref{Fig:Sensing_Power_SINR_Simulation} highlight the sensing-communication trade-off, where higher sensing power enhances accuracy at the cost of increased communication power consumption. Overall, as observed from the numerical results in this section, improved sensing performance can be achieved with longer dwell time, stronger cooperation, and higher sensing power. This improvement supports the sensing feasibility of the proposed frameworks and extends their applicability to various ISAC network environments.}

\section{Conclusion}\label{Sec:Conclusion}
In this paper, we have investigated a downlink multi-satellite ISAC network, where multiple satellites and gateways are deployed to simultaneously provide communication services to ground UEs and perform airborne target sensing. We developed centralized and distributed sensing frameworks based on a practical hybrid analog–digital beamforming architecture. Under the centralized sensing framework, all gateways forward their observations to a CU, and then the collected data are jointly processed to estimate the target positions. Conversely, under the distributed framework, each gateway individually estimates the target positions and forwards its results to the CU. Through simulations, we demonstrated the superiority of the proposed frameworks over existing sensing frameworks while guaranteeing communication performance. We also analyzed the sensing–communication trade-off from the viewpoints of sensing accuracy and communication power consumption under the proposed frameworks.


For further work in this area, a robust design with hardware impairments, such as imperfect synchronization \cite{shi2024joint} and phase noise \cite{dong2025fundamental}, can be explored. \textcolor{black}{An extended evaluation of communication-side metrics in terms of the SINR distribution, ergodic rate \cite{meng2024scaling} and outage probability can also be studied to validate the dual functionality of cooperative multi-satellite ISAC networks.}

\begin{figure*}[!ht]
\vspace{-7mm}
\textcolor{black}{
    \begin{align}
    &\frac{\partial \mathbf{v}^{\textrm{gat-tar}}_{x,l,k}}{\partial[\mathbf{q}^{\textrm{tar}}_{k}]_{b}} = j\pi \left(\sin\phi^{\textrm{gat-tar}}_{l,k}\cos\theta^{\textrm{gat-tar}}_{l,k}\frac{\partial \phi^{\textrm{gat-tar}}_{l,k}}{\partial[\mathbf{q}^{\textrm{tar}}_{k}]_{b}}
    +\cos\phi^{\textrm{gat-tar}}_{l,k}\sin\theta^{\textrm{gat-tar}}_{l,k}\frac{\partial \theta^{\textrm{gat-tar}}_{l,k}}{\partial[\mathbf{q}^{\textrm{tar}}_{k}]_{b}}\right)\textrm{diag}(0,1,\cdots,N_{x}^{\textrm{gat}}-1)\mathbf{v}^{\textrm{gat-tar}}_{x,l,k},\label{array response derivative 1} \tag{69}\\
    &\frac{\partial \mathbf{v}^{\textrm{gat-tar}}_{y,l,k}}{\partial[\mathbf{q}^{\textrm{tar}}_{k}]_{b}} = j\pi\left(-\cos\phi^{\textrm{gat-tar}}_{l,k}\cos\theta^{\textrm{gat-tar}}_{l,k}\frac{\partial \phi^{\textrm{gat-tar}}_{l,k}}{\partial[\mathbf{q}^{\textrm{tar}}_{k}]_{b}}+\sin\phi^{\textrm{gat-tar}}_{l,k}\sin\theta^{\textrm{gat-tar}}_{l,k}\frac{\partial \theta^{\textrm{gat-tar}}_{l,k}}{\partial[\mathbf{q}^{\textrm{tar}}_{k}]_{b}}
    \right)\textrm{diag}(0,1,\cdots,N_{y}^{\textrm{gat}}-1)\mathbf{v}^{\textrm{gat-tar}}_{y,l,k} ,\label{array response derivative 2}\tag{70}\\
    &\frac{\partial \mathbf{v}^{\textrm{sat-tar}}_{x,i,k}}{\partial[\mathbf{q}^{\textrm{tar}}_{k}]_{b}} = j\pi\left(\sin\phi^{\textrm{sat-tar}}_{i,k}\cos\theta^{\textrm{sat-tar}}_{i,k}\frac{\partial \phi^{\textrm{sat-tar}}_{i,k}}{\partial[\mathbf{q}^{\textrm{tar}}_{k}]_{b}}
    +\cos\phi^{\textrm{sat-tar}}_{i,k}\sin\theta^{\textrm{sat-tar}}_{i,k}\frac{\partial \theta^{\textrm{sat-tar}}_{i,k}}{\partial[\mathbf{q}^{\textrm{tar}}_{k}]_{b}}\right)\textrm{diag}(0,1,\cdots,N_{x}^{\textrm{sat}}-1)\mathbf{v}^{\textrm{sat-tar}}_{x,i,k},\label{array response derivative 3}\tag{71}\\
    &\frac{\partial \mathbf{v}^{\textrm{sat-tar}}_{y,i,k}}{\partial[\mathbf{q}^{\textrm{tar}}_{k}]_{b}} = j\pi\left(-\cos\phi^{\textrm{sat-tar}}_{i,k}\cos\theta^{\textrm{sat-tar}}_{i,k}\frac{\partial \phi^{\textrm{sat-tar}}_{i,k}}{\partial[\mathbf{q}^{\textrm{tar}}_{k}]_{b}}+\sin\phi^{\textrm{sat-tar}}_{i,k}\sin\theta^{\textrm{sat-tar}}_{i,k}\frac{\partial \theta^{\textrm{sat-tar}}_{i,k}}{\partial[\mathbf{q}^{\textrm{tar}}_{k}]_{b}}
    \right)\textrm{diag}(0,1,\cdots,N_{y}^{\textrm{sat}}-1)\mathbf{v}^{\textrm{sat-tar}}_{y,i,k}.\label{array response derivative 4}\tag{72}
    \end{align}
    \hrulefill
}
\vspace{-4mm}
\end{figure*} 

\vspace{-2mm}
\appendices
\section{Derivation of the FIM}
\textcolor{black}{We first calculate the partial derivatives of $\mu_{l}(t)$ with respect to the unknown parameters as
\begin{align}
&\frac{\partial \mu_{l}(t)}{\partial [\mathbf{q}^{\textrm{tar}}_{k}]_{b}} = \sum_{i\in\mathcal{I}}\rho_{i,k,l}\frac{\partial a_{i,k,l}^{\textrm{gat}}(t)}{\partial[\mathbf{q}^{\textrm{tar}}_{k}]_{b}},~ b\in\{1,2,3\}\label{partial derivative 1}\\
&\frac{\partial \mu_{l}(t)}{\partial \textrm{Re}(\rho_{i,k,l'})} = a_{i,k,l}^{\textrm{gat}}(t)\delta_{l,l'},~i\in\mathcal{I},\,k\in\mathcal{K},\,l,l'\in\mathcal{L},\label{partial derivative 2}\\
&\frac{\partial \mu_{l}(t)}{\partial \textrm{Im}(\rho_{i,k,l'})} = ja_{i,k,l}^{\textrm{gat}}(t)\delta_{l,l'},~i\in\mathcal{I},\,k\in\mathcal{K},\,l,l'\in\mathcal{L},\label{partial derivative 3}
\end{align}
where $\delta_{l,l'}$ is the Kronecker delta. By substituting \eqref{partial derivative 1}-\eqref{partial derivative 3} into \eqref{FIM}, the block elements of $\mathbf{J}_{\boldsymbol{\eta}}$ are rewritten as
\begin{align}
&[\mathbf{J}_{\boldsymbol{\vartheta}\boldsymbol{\vartheta}}]_{3(k-1)+b,3(k'-1)+b'} 
= 2\sum_{l\in\mathcal{L}}\sum_{t\in\mathcal{T}}\frac{1}{\zeta_l^2}\nonumber\\
&\!\times\!\textrm{Re}\bigg(\Big(\sum_{i\in\mathcal{I}}\rho_{i,k,l}\frac{\partial a_{i,k,l}^{\textrm{gat}}(t)}{\partial[\mathbf{q}^{\textrm{tar}}_{k}]_{b}}\Big)^H
 \Big(\sum_{i\in\mathcal{I}}\rho_{i,k',l}\frac{\partial a_{i,k',l}^{\textrm{gat}}(t)}{\partial[\mathbf{q}^{\textrm{tar}}_{k'}]_{b'}}\Big)\bigg),\!\label{block FIM}\\
&[\mathbf{J}_{\boldsymbol{\vartheta}\boldsymbol{\rho}_\textrm{R}}]_{3(k-1)+b,k'+(i'-1)K+(l'-1)IK} \nonumber\\
&= 2\sum_{t\in\mathcal{T}}\frac{1}{\zeta_{l'}^2}
\textrm{Re}\bigg(\Big(\sum_{i\in\mathcal{I}}\rho_{i,k,l'}\frac{\partial a_{i,k,l'}^{\textrm{gat}}(t)}{\partial[\mathbf{q}^{\textrm{tar}}_{k}]_{b}}\Big)^H a_{i',k',l'}^{\textrm{gat}}(t)\bigg),\\
&[\mathbf{J}_{\boldsymbol{\vartheta}\boldsymbol{\rho}_\textrm{I}}]_{3(k-1)+b,k'+(i'-1)K+(l'-1)IK} \nonumber\\
&= 2\sum_{t\in\mathcal{T}}\frac{1}{\zeta_{l'}^2}
\textrm{Re}\bigg(\Big(\sum_{i\in\mathcal{I}}\rho_{i,k,l'}\frac{\partial a_{i,k,l'}^{\textrm{gat}}(t)}{\partial[\mathbf{q}^{\textrm{tar}}_{k}]_{b}}\Big)^H ja_{i',k',l'}^{\textrm{gat}}(t)\bigg),\\
&[\mathbf{J}_{\boldsymbol{\rho}_\textrm{R}\boldsymbol{\rho}_\textrm{R}}]_{k+(i-1)K+(l-1)IK,k'+(i'-1)K+(l'-1)IK} \nonumber\\
&=[\mathbf{J}_{\boldsymbol{\rho}_\textrm{I}\boldsymbol{\rho}_\textrm{I}}]_{k+(i-1)K+(l-1)IK,k'+(i'-1)K+(l'-1)IK}\nonumber\\ 
&= 2\sum_{t\in\mathcal{T}}\frac{1}{\zeta_l^2}
\textrm{Re}\bigg(a_{i,k,l}^{\textrm{gat}}(t)^H a_{i',k',l}^{\textrm{gat}}(t)\bigg)\delta_{l,l'},\\
&[\mathbf{J}_{\boldsymbol{\rho}_\textrm{R}\boldsymbol{\rho}_\textrm{I}}]_{k+(i-1)K+(l-1)IK,k'+(i'-1)K+(l'-1)IK} \nonumber\\
&= 2\sum_{t\in\mathcal{T}}\frac{1}{\zeta_l^2}
\textrm{Re}\bigg(a_{i,k,l}^{\textrm{gat}}(t)^H ja_{i',k',l}^{\textrm{gat}}(t)\bigg)\delta_{l,l'}.
\end{align}
From the definition of $\boldsymbol{\xi}$, we have $\mathbf{J}_{\boldsymbol{\vartheta}\boldsymbol{\xi}} = [\mathbf{J}_{\boldsymbol{\vartheta}\boldsymbol{\rho}_\textrm{R}}, \mathbf{J}_{\boldsymbol{\vartheta}\boldsymbol{\rho}_\textrm{I}}]$ and
\begin{align}
\mathbf{J}_{\boldsymbol{\xi}\boldsymbol{\xi}}=
\begin{bmatrix}
\mathbf{J}_{\boldsymbol{\rho}_\textrm{R}\boldsymbol{\rho}_\textrm{R}} & \mathbf{J}_{\boldsymbol{\rho}_\textrm{R}\boldsymbol{\rho}_\textrm{I}}\\
\mathbf{J}_{\boldsymbol{\rho}_\textrm{R}\boldsymbol{\rho}_\textrm{I}}^T & \mathbf{J}_{\boldsymbol{\rho}_\textrm{I}\boldsymbol{\rho}_\textrm{I}} 
\end{bmatrix}.
\end{align}
We now calculate the derivative of $a_{i,k,l}^{\textrm{gat}}(t)$ with respect to the target position. Specifically, by applying the product rule, the derivative is obtained as
\begin{align}
\frac{\partial a_{i,k,l}^{\textrm{gat}}(t)}{\partial[\mathbf{q}^{\textrm{tar}}_{k}]_{b}}&=
\frac{\partial\beta_{i,k,l}}{\partial[\mathbf{q}^{\textrm{tar}}_{k}]_{b}}\varphi_{l,k}(t)\varpi_{i,k}(t)+ \beta_{i,k,l}\frac{\partial\varphi_{l,k}(t)}{\partial[\mathbf{q}^{\textrm{tar}}_{k}]_{b}}\varpi_{i,k}(t)
\nonumber\\
&~~~+ \beta_{i,k,l}\varphi_{l,k}(t)\frac{\partial\varpi_{i,k}(t)}{\partial[\mathbf{q}^{\textrm{tar}}_{k}]_{b}},
\end{align}
where $\beta_{i,k,l}$, $\varphi_{l,k}(t)$, and $\varpi_{i,k}(t)$ are defined as
\begin{align}
&\beta_{i,k,l} = \sqrt{\frac{\lambda^2 A^{\textrm{gat}}_{\textrm{Rx}}A^{\textrm{sat}}_{\textrm{Tx}}}{64\pi^3 (d^{\textrm{gat-tar}}_{l,k})^2(d^{\textrm{sat-tar}}_{i,k})^2}} e^{-j2\pi\frac{ d^{\textrm{gat-tar}}_{l,k}+d^{\textrm{sat-tar}}_{i,k}}{\lambda}},\\
&\varphi_{l,k}(t) = \mathbf{w}^{H}_{l}(t) \mathbf{v}^{\textrm{gat-tar}}_{l,k},   ~~\varpi_{i,k}(t) = (\mathbf{v}^{\textrm{sat-tar}}_{i,k})^H\mathbf{x}_{i}(t).
\end{align}
Next, the partial derivative of $\beta_{i,k,l}$ with respect to $[\mathbf{q}^{\textrm{tar}}_{k}]_{b}$ is 
\begin{align}
\frac{\partial \beta_{i,k,l}}{\partial[\mathbf{q}^{\textrm{tar}}_{k}]_{b}} &= \beta_{i,k,l}\Bigg[\frac{[\mathbf{q}^{\textrm{gat}}_{l}]_b-[\mathbf{q}^{\textrm{tar}}_{k}]_b}{(d^{\textrm{gat-tar}}_{l,k})^2} + \frac{[\mathbf{q}^{\textrm{sat}}_{i}]_b-[\mathbf{q}^{\textrm{tar}}_{k}]_b}{(d^{\textrm{sat-tar}}_{i,k})^2} \nonumber\\
&~+j\frac{2\pi}{\lambda}\Bigg(\frac{[\mathbf{q}^{\textrm{gat}}_{l}]_b-[\mathbf{q}^{\textrm{tar}}_{k}]_b}{d^{\textrm{gat-tar}}_{l,k}} + \frac{[\mathbf{q}^{\textrm{sat}}_{i}]_b-[\mathbf{q}^{\textrm{tar}}_{k}]_b}{d^{\textrm{sat-tar}}_{i,k}}\Bigg)\Bigg].
\end{align}
Similarly, the other two derivatives can be obtained as
\begin{align}
&\frac{\partial \varphi_{l,k}(t)}{\partial[\mathbf{q}^{\textrm{tar}}_{k}]_{b}}\!=\!\mathbf{w}^{H}_{l}(t)\!\left(\!\frac{\partial \mathbf{v}^{\textrm{gat-tar}}_{x,l,k}}{\partial[\mathbf{q}^{\textrm{tar}}_{k}]_{b}}\!\otimes\!\mathbf{v}^{\textrm{gat-tar}}_{y,l,k}\!+\!\mathbf{v}^{\textrm{gat-tar}}_{x,l,k}\!\otimes\!\frac{\partial \mathbf{v}^{\textrm{gat-tar}}_{y,l,k}}{\partial[\mathbf{q}^{\textrm{tar}}_{k}]_{b}}\!\right)\!,\!\!\\
&\frac{\partial \varpi_{i,k}(t)}{\partial[\mathbf{q}^{\textrm{tar}}_{k}]_{b}}\!=\!\left(\!\frac{\partial \mathbf{v}^{\textrm{sat-tar}}_{x,i,k}}{\partial[\mathbf{q}^{\textrm{tar}}_{k}]_{b}}\!\otimes\!\mathbf{v}^{\textrm{sat-tar}}_{y,i,k}\!+\!\mathbf{v}^{\textrm{sat-tar}}_{x,i,k}\!\otimes\!\frac{\partial \mathbf{v}^{\textrm{sat-tar}}_{y,i,k}}{\partial[\mathbf{q}^{\textrm{tar}}_{k}]_{b}}\!\right)^H\!\mathbf{x}_{i}(t).\!\!
\end{align}
Here, the derivatives of the $x$- and $y$-axis array response vectors are given in \eqref{array response derivative 1}-\eqref{array response derivative 4} at the top of this page. The angle derivatives for the gateway-to-target link are given by
\setcounter{equation}{72}
\begin{align}
&\frac{\partial \phi^{\textrm{gat-tar}}_{l,k}}{\partial x^{\textrm{tar}}_{k}} = \frac{\Delta y^{\textrm{gat-tar}}_{l,k}}{(r^{\textrm{gat-tar}}_{l,k})^2}, ~~ \frac{\partial \theta^{\textrm{gat-tar}}_{l,k}}{\partial x^{\textrm{tar}}_{k}} = \frac{\Delta x^{\textrm{gat-tar}}_{l,k}\Delta z^{\textrm{gat-tar}}_{l,k}}{r^{\textrm{gat-tar}}_{l,k}(d^{\textrm{gat-tar}}_{l,k})^2},\\
&\frac{\partial \phi^{\textrm{gat-tar}}_{l,k}}{\partial y^{\textrm{tar}}_{k}} = -\frac{\Delta x^{\textrm{gat-tar}}_{l,k}}{(r^{\textrm{gat-tar}}_{l,k})^2}, ~~ \frac{\partial \theta^{\textrm{gat-tar}}_{l,k}}{\partial y^{\textrm{tar}}_{k}} = \frac{\Delta y^{\textrm{gat-tar}}_{l,k}\Delta z^{\textrm{gat-tar}}_{l,k}}{r^{\textrm{gat-tar}}_{l,k}(d^{\textrm{gat-tar}}_{l,k})^2},\\
&\frac{\partial \phi^{\textrm{gat-tar}}_{l,k}}{\partial z^{\textrm{tar}}_{k}} = 0, ~~ \frac{\partial \theta^{\textrm{gat-tar}}_{l,k}}{\partial z^{\textrm{tar}}_{k}} = -\frac{r^{\textrm{gat-tar}}_{l,k}}{(d^{\textrm{gat-tar}}_{l,k})^2},
\end{align}
with $\Delta x^{\textrm{gat-tar}}_{l,k} = x^{\textrm{gat}}_{l}-x^{\textrm{tar}}_{k}$, $\Delta y^{\textrm{gat-tar}}_{l,k} = y^{\textrm{gat}}_{l}-y^{\textrm{tar}}_{k}$, $\Delta z^{\textrm{gat-tar}}_{l,k} = z^{\textrm{gat}}_{l}-z^{\textrm{tar}}_{k}$, and $r^{\textrm{gat-tar}}_{l,k} = \sqrt{(\Delta x^{\textrm{gat-tar}}_{l,k})^2 +(\Delta y^{\textrm{gat-tar}}_{l,k})^2}$.
Similarly, for the satellite-to-target link, we have
\begin{align}
&\frac{\partial \phi^{\textrm{sat-tar}}_{i,k}}{\partial x^{\textrm{tar}}_{k}} = \frac{\Delta y^{\textrm{sat-tar}}_{i,k}}{(r^{\textrm{sat-tar}}_{i,k})^2}, ~~ \frac{\partial \theta^{\textrm{sat-tar}}_{i,k}}{\partial x^{\textrm{tar}}_{k}} = -\frac{\Delta x^{\textrm{sat-tar}}_{i,k}\Delta z^{\textrm{sat-tar}}_{i,k}}{r^{\textrm{sat-tar}}_{i,k}(d^{\textrm{sat-tar}}_{i,k})^2},\\
&\frac{\partial \phi^{\textrm{sat-tar}}_{i,k}}{\partial y^{\textrm{tar}}_{k}} = -\frac{\Delta x^{\textrm{sat-tar}}_{i,k}}{(r^{\textrm{sat-tar}}_{i,k})^2}, ~~ \frac{\partial \theta^{\textrm{sat-tar}}_{i,k}}{\partial y^{\textrm{tar}}_{k}} = -\frac{\Delta y^{\textrm{sat-tar}}_{i,k}\Delta z^{\textrm{sat-tar}}_{i,k}}{r^{\textrm{sat-tar}}_{i,k}(d^{\textrm{sat-tar}}_{i,k})^2},\\
&\frac{\partial \phi^{\textrm{sat-tar}}_{i,k}}{\partial z^{\textrm{tar}}_{k}} = 0, ~~ \frac{\partial \theta^{\textrm{sat-tar}}_{i,k}}{\partial z^{\textrm{tar}}_{k}} = \frac{r^{\textrm{sat-tar}}_{i,k}}{(d^{\textrm{sat-tar}}_{i,k})^2}, \label{angle derivatives}
\end{align}
with $\Delta x^{\textrm{sat-tar}}_{i,k} = x^{\textrm{sat}}_{i}-x^{\textrm{tar}}_{k}$, $\Delta y^{\textrm{sat-tar}}_{i,k} = y^{\textrm{sat}}_{i}-y^{\textrm{tar}}_{k}$, $\Delta z^{\textrm{sat-tar}}_{i,k} = z^{\textrm{sat}}_{i}-z^{\textrm{tar}}_{k}$, and $r^{\textrm{sat-tar}}_{i,k} = \sqrt{(\Delta x^{\textrm{sat-tar}}_{i,k})^2 +(\Delta y^{\textrm{sat-tar}}_{i,k})^2}$. As a result, the FIM is explicitly obtained by plugging \eqref{block FIM}-\eqref{angle derivatives} into \eqref{FIM matrix}.
}

\vspace{-3mm}
\bibliographystyle{IEEEtran}
\bibliography{ref_TWC}
    
\end{document}